\documentclass[12pt]{article}
\usepackage{cite}
\usepackage{graphics}

\advance\textheight by \topskip
\def\lsim{\mathrel{\lower2.5pt\vbox{\lineskip=0pt\baselineskip=0pt
\hbox{$<$}\hbox{$\sim$}}}}
\def\gsim{\mathrel{\lower2.5pt\vbox{\lineskip=0pt\baselineskip=0pt
\hbox{$>$}\hbox{$\sim$}}}}
\newcommand{\myss}{\mbox{\ss}}
\newcommand{\myL}{\mbox{\L}}

\newcommand{\figuno}{
\begin{figure}[tb]
\begin{center}
\includegraphics{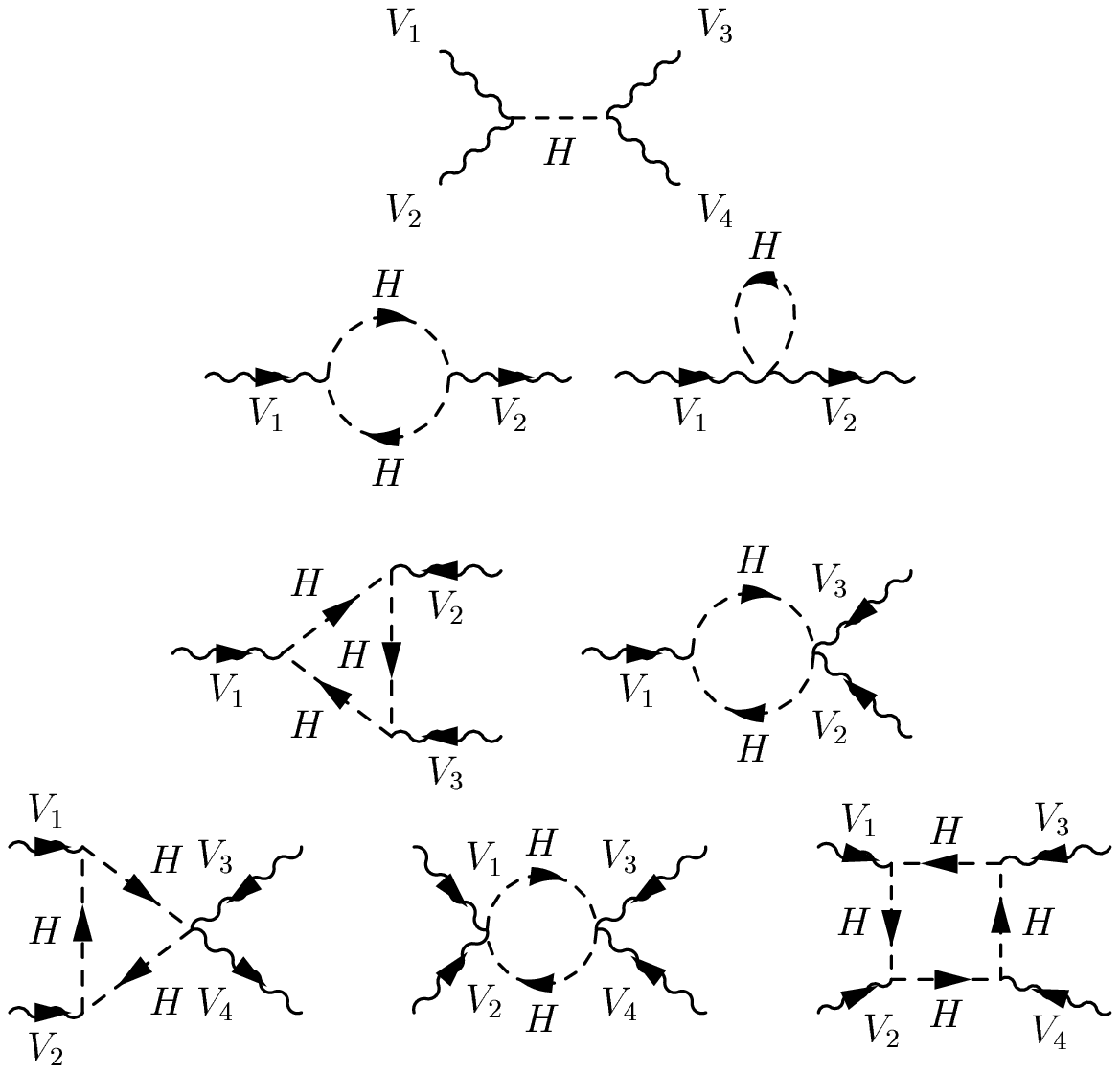}
\end{center}
\caption{Generic Feynman diagrams corresponding to the tree-level 
  and one-loop contributions to the two-, three- and four-point functions of
  electroweak gauge bosons.}
\label{fig:uno}
\end{figure}
}

\begin{document}
\newcommand{\sq}{\tilde{q}}
\newcommand{\sle}{\tilde{l}}
\newcommand{\sneut}{\tilde{\nu}}
\newcommand{\cpm}{\tilde{\chi}^{\pm}}
\newcommand{\neut}{\tilde{\chi}^{o}}
\newcommand{\hsm}{H_{SM}}
\newcommand{\hz}{h^o}
\newcommand{\Hz}{H^o}
\newcommand{\Az}{A^o}
\newcommand{\hpm}{H^\pm}
\newcommand{\hSM}{H_{\scriptstyle SM}}
\newcommand{\mhSM}{M_{H_{\scriptscriptstyle SM}}}
\newcommand{\mw}{m_W}
\newcommand{\mzns}{m_Z}
\newcommand{\mew}{M_{EW}}
\newcommand{\ma}{m_{A^o}}
\newcommand{\mh}{m_{h^o}}
\newcommand{\mH}{m_{H^o}}
\newcommand{\mhp}{m_{H^\pm}}
\newcommand{\msto}{\tilde m_{{t}_1}}
\newcommand{\mstt}{\tilde m_{{t}_2}}
\newcommand{\msbo}{\tilde m_{{b}_1}}
\newcommand{\msbt}{\tilde m_{{b}_2}}
\def\gs{SU(2)_{\rm L} \times U(1)_{\rm Y}}
\def\w{{\tilde{W}^{\pm}}}
\def\h{{\tilde{H}^{\pm}}}
\def\sw{s_{{\scriptscriptstyle W}}}
\def\cw{c_{{\scriptscriptstyle W}}}
\def\mi{{\tilde{M}^2}_{{\scriptscriptstyle i}}}
\def\mj{{\tilde{M}^2}_{{\scriptscriptstyle j}}}
\newcommand{\dx}{dx} 
\newcommand{\dy}{dy} 
\newcommand{\slas}[1]{\rlap/ #1}
\newcommand{\diag}{{\rm diag}}
\newcommand{\Tr}{{\rm Tr}}
\newcommand{\hi}{{H}}
\def\at{{A_{\scriptscriptstyle t}}}
\def\ab{{A_{\scriptscriptstyle b}}}
\def\qt{{Q_{\scriptscriptstyle t}}}
\def\qb{{Q_{\scriptscriptstyle b}}}
\newcommand{\sfe}{{\tilde f}}
\newcommand{\sg}{{\tilde \chi}}
\newcommand{\sne}{{\tilde \chi^o}}
\newcommand{\scp}{{\tilde \chi^+}}
\def\mz{{m}_{\scriptscriptstyle Z}^{2}}
\def\sw{s_{{\scriptscriptstyle W}}}
\def\cw{c_{{\scriptscriptstyle W}}}
\def\cdw{c_{{\scriptscriptstyle 2W}}}
\def\mew{M_{{\scriptscriptstyle EW}}}
\newcommand{\sab}{s_{\alpha\beta}}
\newcommand{\Jslnm}{J_{\sigma\lambda\nu\mu}}
\newcommand{\Jlsnm}{J_{\lambda\sigma\nu\mu}}
\newcommand{\Jslmn}{J_{\sigma\lambda\mu\nu}}
\newcommand{\Jlsmn}{J_{\lambda\sigma\mu\nu}}
\newcommand{\Jmnsl}{J_{\mu\nu\sigma\lambda}}
\newcommand{\Jmnls}{J_{\mu\nu\lambda\sigma}}
\newcommand{\Jnmsl}{J_{\nu\mu\sigma\lambda}}
\newcommand{\Jnmls}{J_{\nu\mu\lambda\sigma}}
\newcommand{\Jsnml}{J_{\sigma\nu\mu\lambda}}
\newcommand{\Jlnms}{J_{\lambda\nu\mu\sigma}}
\newcommand{\Jsmnl}{J_{\sigma\mu\nu\lambda}}
\newcommand{\Jlmns}{J_{\lambda\mu\nu\sigma}}
\newcommand{\Jmlsn}{J_{\mu\lambda\sigma\nu}}
\newcommand{\Jmsln}{J_{\mu\sigma\lambda\nu}}
\newcommand{\Jnlsm}{J_{\nu\lambda\sigma\mu}}
\newcommand{\Jnslm}{J_{\nu\sigma\lambda\mu}}
\newcommand{\Jnsml}{J_{\nu\sigma\mu\lambda}}
\newcommand{\Jnlms}{J_{\nu\lambda\mu\sigma}}
\newcommand{\Jlnsm}{J_{\lambda\nu\sigma\mu}}
\newcommand{\Jsnlm}{J_{\sigma\nu\lambda\mu}}
\newcommand{\Jmsnl}{J_{\mu\sigma\nu\lambda}}
\newcommand{\Jmlns}{J_{\mu\lambda\nu\sigma}}
\newcommand{\Jlmsn}{J_{\lambda\mu\sigma\nu}}
\newcommand{\Jsmln}{J_{\sigma\mu\lambda\nu}}

\vspace*{-1cm}
\begin{flushright}
{\large FTUAM 99/20}\\
{\large hep-ph/0002134}\\
{Feb.-2000}\\
\end{flushright}
\begin{center}
\begin{large}
\begin{bf} THE HIGGS SECTOR OF THE MSSM\\ IN THE DECOUPLING LIMIT\\
\end{bf}
\end{large}
\vspace{0.2cm}
ANTONIO DOBADO $^{\star}$\\
\vspace{0.05cm} 
{\em  Departamento de F{\'\i}sica Te{\'o}rica\\
  Universidad Complutense de Madrid\\
 28040-- Madrid,\ \ Spain} \\
\vspace{0.3cm} 
 MARIA J. HERRERO $^{\dagger}$\\
\vspace{0.05cm}
and\\
\vspace{0.1cm}
SIANNAH PE{\~N}ARANDA $^{\ddag}$\\
\vspace{0.06cm}
{\em  Departamento de F{\'\i}sica Te{\'o}rica\\
  Universidad Aut{\'o}noma de Madrid\\
  Cantoblanco,\ \ 28049-- Madrid,\ \ Spain}
\end{center}
\begin{center}
{\bf ABSTRACT}
\end{center}
\begin{quotation}
\noindent
We study the heavy Higgs sector of the MSSM composed of the 
$H^{\pm}$, $H^0$ and $A^0$ particles 
in the so-called decoupling limit where $m_{A^0}\gg m_Z$. 
By integrating out these 
heavy Higgs particles to
one-loop, we compute the effective action for the electroweak gauge bosons and
find out that, in the decoupling limit, all the heavy Higgs effects can be
absorbed into redefinitions of the Standard Model electroweak parameters. This 
demonstrates explicitely that the decoupling theorem works for the heavy MSSM 
Higgs particles. This is also compared with the paradigmatic and different 
case of the Standard Model heavy Higgs particle. Finally, this work together 
with our two
previous works, complete the demonstration that all the non-standard
particles in the MSSM, namely, squarks, sleptons, charginos, neutralinos and the
heavy Higgs particles, decouple to one-loop from the low energy electroweak 
gauge boson physics. 
\end{quotation}
${\star}$ {\em e-mail: dobado@eucmax.sim.ucm.es}\\
${\dagger}$ {\em e-mail: herrero@delta.ft.uam.es}\\
${\ddag}$ {\em e-mail: siannah@delta.ft.uam.es}\\
\newpage

\section{Introduction}
The absence of any signal from Supersymmetric (SUSY) particles in the existing
data indicates that either SUSY theories are not the proper ones for low energy
physics beyond the Standard Model (SM) or the SUSY spectrum is above the
available energies at present experiments. In the simplest SUSY theory, the
Minimal Supersymmetric Standard Model (MSSM), the predicted spectrum is composed
of squarks $\sq$ and sleptons $\sle$, $\sneut$ for the three generations,
charginos $\cpm_{1,2}$, neutralinos $\neut_{1,2,3,4}$, gluinos $\tilde{g}$, and
the Higgs sector with five Higgs particles, two CP-even Higgs bosons $\hz$ and
$\Hz$, a CP-odd or pseudoscalar Higgs boson $\Az$, and two charged Higgs
particles $\hpm$. Although the precise mass bound varies for each particle, it
is clear that, at present time, there is little room for light MSSM particles,
say 
lighter
than the $W$ gauge boson mass $\mw$. Particularly stringent are the bounds for 
the strongly interacting
particles, the squarks and gluinos with a lower mass limit already above
$200\,GeV$\cite{Caso:1998tx}. Under these circumstances it is a reasonable
hypothesis to think of a mass gap between the SM particles and the
genuine MSSM particles.
In case this energy separation occurs, its size should not be larger than  about
$1\,TeV$, if the MSSM is required to repair the hierarchy problem. We will assume
here the extreme but plausible situation where all the MSSM spectrum lay well
above the electroweak scale $\mew$. 
For the purpose of this paper we just need to assume the existence of this
sizeable gap, but the particular value of the gap width is not relevant. There
is just one exception in this large SUSY mass assumption, the lightest
CP-even $\hz$ particle which stays close to the SM spectrum. It is well known
that when the pseudoscalar mass $\ma$ is very large, that is much larger than
the $Z$ boson mass $\ma\gg\mzns$, the heavy
CP-even, CP-odd and charged Higgs bosons are nearly degenerate,
$\mH\simeq\mhp\simeq\ma$, while the $\hz$ particle reaches its maximal mass
value which, at tree level, is bounded from above by $\mzns$, and when radiative
corrections are included, this upper bound is shifted towards 
$\sim 130\,GeV$~\cite{Okada:1991vk,Haber:1991aw,Barbieri:1991ja,Ellis:1991zd,
Ellis:1991nz,Hempfling:1994qq,Casas:1995us,Haber:1996fp,Quiros:1998bz,Casas:1998vh,Heinemeyer:1998jw,Heinemeyer:1998kz,Espinosa:1999zm}. 
In this so-called decoupling 
limit~\cite{Haber:1993pv}, the lightest SUSY Higgs
boson $\hz$ and the SM Higgs boson $\hsm$ have very similar properties, since
both have similar couplings to fermions and vector bosons and
therefore the task of discriminating between these two particles will be quite
hard. This equality of couplings is exact at tree level when the decoupling
limit is reached asymptotically and both their production rates and decay
branching rations are identical. However, 
it is not known with complete generality if this equality remains beyond
tree level. It is a very interesting
subject, since in case it does not happen it will provide the clue for
discriminating between the SM and MSSM, even in the extreme situation mentioned
above where all the rest of the MSSM spectrum is well above the electroweak
scale and hence not reachable at present experiments. This topic has been
studied by several 
authors~\cite{Drees:1990dx,Gosdzinsky:1991ga,Haber:1993pv,Chankowski:1994eu,Garcia:1994sb,Dabelstein:1995hb,Garcia:1995wu,Garcia:1995wv,Coarasa:1996qa,Djouadi:1998pb,Carena:1996wu,Chankowski:1994er,Alam:2000cs,Inami:1992rb,Haber:1992cn,Coarasa:1996yg,Carena:1999bh}
by looking to particular
observables of interest in phenomenology, as for instance, the parameters $S$,
$T$ and $U$\footnote{or equivalently $\Delta r$, $\Delta\rho$, 
$\Delta\kappa$ or the $\epsilon_i$ parameters.} that measure the radiative 
corrections at LEP~\cite{Haber:1993pv,Inami:1992rb,Haber:1992cn}, the
$\hz$ production rates at LEP and LHC and the decay branching rations of $\hz$
to $\gamma\gamma$~\cite{Djouadi:1998pb} and to 
$f\bar{f}$~\cite{Dabelstein:1995hb,Coarasa:1996yg,Carena:1999bh}. 
Most of these studies analyzed the decoupling of SUSY particles numerically.
Although the numerical analysis are
complicated since they depend on many MSSM parameters, there are indications from
these studies that the SUSY particles indeed tend to  decouple in
the previous observables 
when the SUSY masses are taken numerically very large. In particular, the MSSM
$\hz$ couplings to $\gamma\gamma$\cite{Djouadi:1998pb} and to
$f\bar{f}$\cite{Coarasa:1996yg} seem to approach those of the
$\hsm$ particle in the decoupling limit and in the one loop approximation,
confirming therefore the enormous challenge that will be discriminating between
these two particles at future high energy colliders as the LHC.

In this paper we study the MSSM Higgs sector in the decoupling limit at a more
formal level. Our object of interest is the effective action for the SM
particles and the contributions to this action from the loops of the MSSM
Higgs sector in the limit where all the Higgs particles, except $\hz$, are very
heavy, namely, when $\ma\gg\mzns$. We want to demonstrate the decoupling of the
MSSM Higgs particles {\em {\'a} la 
Appelquist Carazzone}~\cite{Appelquist:1975tg},
meaning that the required 
proof should show that the decoupling theorem also applies for this particular
case. This is the third work belonging to a program that we initiated
in~\cite{Dobado:1997up,Dobado:1999cz} which aims to demonstrate the decoupling of
SUSY particles beyond tree level in each of the MSSM sectors. In generic words,
and by following the Appelquist-Carazzone approach, the
proof of decoupling of SUSY particles at low energies amounts to first compute
the effective action $\Gamma_{\rm eff}[\phi]$ for the SM particles $\phi$
($\phi=q,l,\nu,Z,W^\pm,\gamma,g,\hsm$) that is generated through functional
integration of all the non-standard particles of the MSSM $\tilde\phi$
($\tilde\phi=\sq,\sle,\sneut,\cpm,\neut,\tilde{g},\hpm,\Hz,\Az$)
\begin{equation}
\label{eq:gammaeff}
{\rm e}^{i\Gamma_{eff}[\phi]}=\int [{\rm d}\tilde\phi]\,{\rm e}^{i \Gamma_{\rm M
SSM}
  [\phi,\tilde\phi]}\,\,,
\end{equation}
with
\begin{equation}
\label{eq:gammaeffmssm}
\Gamma_{\rm MSSM}[\phi,\tilde\phi] \equiv \int \dx{\cal L}_{\rm
  MSSM}(\phi,\tilde \phi)\,\,;\,\,{\rm d}x\equiv{\rm d}^4x \,\,,
\end{equation}
and ${\cal L}_{\rm MSSM}$ is the MSSM Lagrangian.

Secondly, one must perform a large SUSY mass expansion of 
$\Gamma_{\rm  eff}[\phi]$ to  be valid for low energies, say 
$\mew\ll M_{\tilde {\phi}}$, and, as a result, one should get finally the
following behaviour,
\begin{equation}
\label{eq:gammaefflimit}
\Gamma_{\rm eff}[\phi] = \hat{\Gamma}_{\rm SM}[\phi]+
{\cal O}\left[\left(\frac{\mew}{M_{\tilde{\phi}}}\right)^n\right]\,\,,
\end{equation}
which means that all the effects of the heavy SUSY particles $\tilde\phi$ can be
absorbed into redefinitions of the SM couplings and wave functions of the SM
fields $\phi$, or else they are suppressed by inverse powers of the heavy masses
$M_{\tilde\phi}$ and therefore vanish in the asymptotic limit
$M_{\tilde\phi}\to\infty$.  We believe that only an explicit computation as the
one just outlined can be considered as a formal and general proof of decoupling
of non-standard particles from the low energy SM physics.

We have started this program with the computation of the part of the effective
action for the electroweak gauge bosons, but, of course, a complete proof of
decoupling will require to obtain the total effective action for the other SM
particles as well, namely, the fermions, the gluon and the SM Higgs particle
itself. In particular, the study of the $h^{0}b\bar b$ vertex is one of the most
interesting observables in the Higgs phenomenology~\cite{David:1999yo}. The reason to start with the electroweak gauge boson sector is, first,
for simplicity and, second, because we were interested in studing the
implications for some of the precision observables at LEP with external gauge
bosons as the $S$, $T$ and $U$ or related parameters. We have proved that, to one loop level, the functional
integration of the various MSSM sparticle sectors factorize in the effective
action for electroweak bosons and, therefore, this integration can be performed 
sector by
sector separately. In~\cite{Dobado:1997up,Dobado:1999cz} we have completed the
integration of squarks, sleptons, charginos and neutralinos in the MSSM 
to one loop, and have demonstrated their decoupling in the large SUSY masses
limit. Since the asymptotic behaviour of the Feynman loop integrals appearing in
the computation depend on the relative sizes of the various sparticle masses in
the loop propagators, one must perform the computation by assuming a particular
hypothesis for these masses. We assumed
in~\cite{Dobado:1997up,Dobado:1999cz} that the large SUSY masses limit is taken
for each sector such that $\mew^2\ll M_{\tilde \phi_i}^2 \;\forall i$,
 but with
$|M_{\tilde \phi_i}^2-M_{\tilde \phi_j}^2| \ll 
|M_{\tilde \phi_i}^2+M_{\tilde \phi_j}^2|$ 
if $i\not=j$. That is, all the SUSY masses are large as compared to the
electroweak scale but they are close to each other. 
 This is a plausible hypothesis in the MSSM but is not the most
general one for all the sectors. In particular for the squarks of the third
generation where, even assuming a common soft-SUSY-breaking mass, one has
$(\msto^2-\mstt^2) \simeq m_t (\at-\mu \cot\beta)$ and 
$(\msbo^2-\msbt^2)\simeq m_b (\ab-\mu\tan\beta)$ and, therefore, for large enough
values of $\at$, $\ab$, $\mu$ and/or $\tan\beta$ the previous hypothesis may not
hold. In consequence, for these particular cases where 
$|M_{\tilde \phi_i}^2-M_{\tilde \phi_j}^2| \simeq 
{\cal O}|M_{\tilde \phi_i}^2+M_{\tilde \phi_j}^2|$ for $i\not=j$ an
 independent demonstration of decoupling should be done.

In the present work we complete the computation of the effective action for
electroweak gauge bosons to one loop by integrating out the heavy MSSM Higgs
particles, namely the charged $\hpm$, the pseudoscalar $\Az$ and the heaviest
CP-even Higgs boson $\Hz$. We then perform the large mass expansion which in
the Higgs sector case corresponds to work in the above mentioned 
decoupling limit. Notice that for the Higgs sector the
previous assumption for the relative Higgs mass values,
$|m_{H_i}^2-m_{H_j}^2|\ll |m_{H_i}^2+m_{H_j}^2|$ if $i\not=j$ holds trivially,
since when $\ma\gg\mzns$ the four heavy Higgs bosons, $\hpm$, $\Az$ and $\Hz$ tend
to be degenerate with a mass close to $\ma$.

The paper is organized as follows. In the second section we define the effective
action for the electroweak gauge bosons and summarize the relevant part of the MSSM
lagrangian for the purpose of integration of the MSSM Higgs sector to one loop
level. The exact results to one loop of the contributions to the effective
action from the 2, 3, and 4 point electroweak gauge bosons functions are
presented in section three. We also analyze in that section the behaviour of
these functions in the decoupling limit, $\ma\gg\mzns$, and present the
corresponding asymptotic results in terms of the large Higgs masses $\mhp$,
$\ma$, $\mH$. In section four the previous asymptotic expressions are rewritten
in a form that will allow us to conclude on the decoupling of the Higgs sector
{\em {\'a} la Appelquist Carazzone} as announced. In particular, by using the
common language of renormalization, the required
redefinitions of the SM couplings and wave functions for the electroweak bosons
are presented in the form of specific contributions to the SM
counterterms. Section five is devoted to a comparison with the paradigmatic and
dramatically different case of the SM with a very heavy Higgs particle, 
$\mew\ll M_{\hsm}$, which is well known not to decouple from low
energy electroweak physics~\cite{Einhorn:1981cy,Appelquist:1980vg,Longhitano:1980iz,Longhitano:1981tm,Herrero:1994nc,Herrero:1995iu,Dittmaier:1995cr,Espriu:1995rm}. We find
illustrative to perform this comparison in the language 
of the effective action.
This non-decoupling of the SM Higgs particle
has been shown to manifest at one loop level in several
observables, as for instance 
$\Delta \rho$~\cite{Longhitano:1980iz,Longhitano:1981tm,Bohm:1986rj}, 
and it is being very relevant 
in the indirect Higgs searches at the present colliders.
In section five we reobtain this non-decoupling
behavior by computing the effective action for electroweak gauge bosons after
integration to one loop of the SM Higgs particle and by studying its large
$M_{\hsm}$ expansion. We will see that the non-decoupling of the Higgs particle
manifests in this context as a violation of the decoupling theorem in the four
point electroweak gauge functions. Finally, the conclusions of this work are
summarized in section six.

\section{Integration of the MSSM Higgs sector to one loop}
The effective action for the electroweak gauge bosons, $\Gamma_{\rm eff}[V]$
($V=A,Z,W^\pm$) gets contributions to one loop from all the MSSM sectors, 
except from gluinos which will start contributing at and beyond two loops.
 This effective action is defined through functional integration
of all the sfermions $\sfe$ ($\sq,\sle,\sneut$), neutralinos $\neut$
($\neut_{1\ldots4}$), charginos $\cpm$ ($\cpm_{1,2}$), and the Higgs bosons
$\hi$ ($\hpm,\Hz,\Az$)  by:
\begin{equation}
\label{eq:gamameffallsectors}
e^{i \Gamma_{eff} [{\scriptscriptstyle V}]} =
\int [d\tilde{f}] [d\tilde{f}^{*}]  [d\tilde{\chi}^{+}] [d\bar{\tilde{\chi}}^{+}] [d\tilde{\chi}^{o}] [d\hi]
e^{i \Gamma_{\rm MSSM} [{\scriptscriptstyle V},\tilde{f},\tilde{\chi}^{+},\neut,\hi]}\,\,,
\end{equation}
where the relevant part of the MSSM classical action can be written as,
\begin{eqnarray}
\label{eq:gammaMSSM}
\Gamma_{\rm MSSM}[V,\sfe, \scp,\sne,\hi]&\equiv&  
\int\dx {\cal L}_{\rm MSSM}(V,\sfe, \scp,\sne,\hi) \nonumber\\ 
&=& \int\dx {\cal L}^{(0)} (V)+ \int\dx{\cal L}_{\sfe}(V,\sfe)+
\int\dx{\cal L}_{\sg}(V,\sg)+\int\dx{\cal L}_{\hi}(V,\hi) \nonumber\\
&\equiv&\Gamma_0[V]+
\Gamma_{\sfe}[V,\sfe]+\Gamma_{\sg}[V,\sg]+\Gamma_{\hi}[V,\hi]\,\,.
\end{eqnarray}
Here, ${\cal L}^{(0)}(V)$ is the free gauge boson lagrangian at tree level, and 
${\cal  L}_{\sfe}$, ${\cal L}_{\sg}$ and ${\cal L}_{\hi}$ are the lagrangians
of sfermions, {\it inos} (i.e.\ charginos and neutralinos) and Higgs bosons
respectively. By looking into the 
particular form of these lagrangians it is inmediate to see that the 
integration of the various sectors at the one-loop level can be factorized out,
 and their contributions to the
effective action can be computed separately sector by sector.

In~\cite{Dobado:1997up,Dobado:1999cz} we have performed the complete integration
to one loop of the sfermions and {\it inos} sectors. Here we present the corresponding
integration of the heavy Higgs sector defined as,
\begin{equation}
\label{eq:gamameffhi}
e^{i \Gamma_{eff}^{\hi} [{\scriptscriptstyle V}]} =
\int   [d\hi]
e^{i \int\dx \left({\cal L}^{(0)} (V)+ {\cal L}_{\hi}(V,\hi)\right)}\,\,,
\end{equation}
where we have introduced a short hand notation for the 
heavy Higgs particles,   
\begin{equation}
\label{eq:Hpesado}
\hi = \left(
\begin{array}{l}
H^1 \\  H^2 \\
H^{o} \\  A^{o} 
\end{array}
\right)\,, 
\end{equation}
with $H^1$ and $H^2$ being related to the physical charged Higgs particles by
$
\hpm\equiv\frac{1}{\sqrt{2}}\left(H^1 \pm i H^2\right)
$,
and ${\cal L}_{\hi}(V,\hi)$ is the relevant MSSM Higgs sector lagrangian that is
given by,
\begin{equation}
\label{eq:lagh}
{\cal L}_{{\hi}} (V, \hi)= {\cal L}^{(0)} (\hi)+
{\cal L}_{\hi V V}+ {\cal L}_{\hi \hi V}+{\cal L}_{\hi \hi V V}\,.
\end{equation}
Here  ${\cal L}^{(0)}(H)$ is the free lagrangian for the heavy Higgs particles,
\begin{equation}
\label{eq:lagharbol}
{\cal L}^{(0)}(\hi)= \frac{1}{2} \left(\partial_{\mu} \hi^{T} 
\partial^{\mu} \hi - \hi^{T} M_{\hi}^{2} \hi  \right)\,, 
\end{equation}
the squared mass matrix is given in terms of the physical Higgs boson masses
by
\begin{equation}
\label{eq:matrixH}
M_{\hi}^{2} \equiv\, \diag (m^{2}_{\scriptstyle {H^{+}}},
m^{2}_{\scriptstyle {H^{+}}}, m^{2}_{\scriptstyle {H^{o}}},
m^{2}_{\scriptstyle {A^{o}}})\,\,,\,\,\,\,
m_{\scriptstyle {H^{+}}}=m_{\scriptstyle {H^{-}}}\,.
\end{equation}
and we have used the superscript $T$ to denote the transpose matrix.
The interaction lagrangian pieces can be written as 
follows~\cite{Gunion:1986yn},
\begin{eqnarray}
\label{eq:lagintH}
{\cal L}_{\hi V V} &=& {\cal B}^T \hi\,,\nonumber\\ 
{\cal L}_{\hi \hi V} &=& \hi^T \vee^{(1) \mu} 
\stackrel{\leftrightarrow}{{\partial}_{\mu}} \hi \,,\nonumber\\ 
{\cal L}_{\hi \hi V V}  &=& \hi^T \vee^{(2)} \hi\,.  
\end{eqnarray}
where
\begin{equation}
\label{eq:betaH}
{\cal B} \equiv \left(
\begin{array}{c}
  0\\ 0\\
 g c_{\alpha\beta}\left({m}_{\scriptscriptstyle W}W_{\mu}^{+}W^{\mu-}+
 \frac{{m}_{\scriptscriptstyle Z}}{2c_W}Z_{\mu}Z^{\mu}\right)\\
 0
\end{array}
\right)\,,
\end{equation}
and  $\vee^{(1) \mu}$, $\vee^{(2)}$ are  the $4\times4$ Higgs interaction
matrices with one and two gauge bosons respectively defined by,
\begin{equation} 
\label{eq:nuevasV} 
\begin{array}{l} 
\displaystyle  
\vee^{(1)^{\mu}} \left\{ 
\begin{array}{l} 
[V^{(1) \mu}]_{ij}= 0\, {\mbox{ if }} i=j\,\,,\,\,\, 
[V^{(1) \mu}]_{ij}=-[V^{(1) \mu}]_{ji}\, {\mbox{ if }} i\neq j\,,\\ 
{[V^{(1) \mu}]}_{12}= eA^{\mu}+\frac{gc_{2W}}{2c_W}Z^{\mu}\,,\,\, 
\displaystyle [V^{(1) \mu}]_{13}=\frac{g}{2}\,s_{\alpha\beta} 
\,W_{2}^{\mu}\,,\,\, 
\displaystyle[V^{(1) \mu}]_{14}=\frac{g}{2}W_{1}^{\mu}\\ 
\displaystyle{[V^{(1) \mu}]}_{23}=-\frac{g}{2}\,s_{\alpha\beta}\, 
W_{1}^{\mu}\,,\,\, 
[V^{(1) \mu}]_{24}=\frac{g}{2}W_{2}^{\mu}\,,\,\, 
[V^{(1) \mu}]_{34}=-\frac{g}{2c_W}s_{\alpha\beta}Z^{\mu} 
\end{array}\right.\nonumber\\ \nonumber\\ 
\displaystyle  
\vee^{(2)} \left\{ 
\begin{array}{l} 
[V^{(2)}]_{ij}=[V^{(2)}]_{ji}\,\forall \, i,j\,,\,\, 
[V^{(2)}]_{12}=[V^{(2)}]_{34}=0\,,\\ 
\displaystyle {[V^{(2)}]}_{11}=[V^{(2)}]_{22}= 
2\left[\frac{g^2}{4}W^{+}_{\mu}W^{\mu-} 
+\frac{g^2\,c_{2\scriptscriptstyle W}^{2}}{8\cw^{2}} 
Z_{\mu}Z^{\mu}+\frac{e^2}{2}A_{\mu}A^{\mu} 
+\frac{eg\,c_{2\scriptscriptstyle W}}{2\cw}A_{\mu}Z^{\mu}\right]\,,\\ 
\displaystyle [V^{(2)}]_{33}=[V^{(2)}]_{44}= \,
2\left[\frac{g^2}{4}W^{+}_{\mu}W^{\mu-}+ 
\frac{g^2}{8\cw^{2}}Z_{\mu}Z^{\mu}\right]\,,\\ 
\displaystyle [V^{(2)}]_{i3}=s_{\beta\alpha}\, 
\left[-\frac{eg}{2}A_{\mu}W^{i\,\mu}+ 
\frac{g^{2}\,\sw^{2}}{2\cw}Z_{\mu}W^{i\,\mu}\right],\,\,i=1,2\,,\\ 
\displaystyle [V^{(2)}]_{i4}=\left[-\frac{eg}{2}A_{\mu}W^{i\,\mu}+ 
\frac{g^{2}\,\sw^{2}}{2\cw}Z_{\mu}W^{i\,\mu}\right],\,\,i=1,2\,. 
\end{array}\right. 
\end{array} 
\end{equation} 
Here, as usual, $g$ and $e$ are the electroweak and
electromagnetic couplings respectively, and we have used a shorthand notation
for the sines and cosines of the weak angle $\theta_W$ and the $\beta$ angle
($\tan\beta\equiv \frac{v_2}{v_1})$) given by
\begin{eqnarray} 
\label{eq:scab}  
s_{\alpha\beta} \equiv \sin (\alpha-\beta)\,&,&\, 
c_{\alpha\beta} \equiv \cos (\alpha-\beta)\,,\nonumber\\ 
c_{2\scriptscriptstyle W} \equiv \cos 2\theta_W\,&,& \,  
s_{2\scriptscriptstyle W} \equiv \sin 2\theta_W\,,\nonumber\\ 
\cw \equiv \cos \theta_W\,&,& \, \sw \equiv \sin \theta_W\,. 
\end{eqnarray} 
Correspondingly, we can define the various contributions to the classical action
by,
\begin{equation}
  \label{eq:gammash}
  \Gamma_{\hi}[V,\hi]=\langle {\cal B}^T\, \hi \rangle + 
 \frac{1}{2}\, \langle\,{\hi}^T\, A_\hi\, \hi\,\rangle \,\,,
\end{equation}
where,
\begin{eqnarray}
\label{eq:notopH}
A_\hi &\equiv & A_\hi^{(0)}+A_\hi^{(1)}+A_\hi^{(2)}\,,\nonumber\\
\langle{\cal B}^T \hi\rangle &\equiv & \int {\rm
  d}\tilde{k}\,{\cal B}^T_k\,\hi_{k}\,,\nonumber\\
\langle\,{\hi}^T\, A_\hi\, \hi\,\rangle  &\equiv & 
 \int {\rm d}\tilde{k}\, {\rm d}\tilde{p}\,{\hi}^{T}_k A_{\hi k
   p}^{(i)}\, {\hi}_p\,,\,\,\,i=0,1,2\,.
\end{eqnarray}
with,
\begin{equation}
\label{eq:difmom}
{\rm d}\tilde{k} \equiv \frac{d^4k}{(2\pi)^4}\,\,,
\end{equation}
and we have chosen the representation in momentum space which is more
convenient for functional integration,
\begin{eqnarray}
  \label{eq:opersHdef}
  A_{\hi kp}^{(0)}&\equiv& (2\pi)^{4}\,\delta(k+p)\, 
(k^2-M_{\hi}^2)\,,\nonumber\\
  A_{\hi kp}^{(1)}&\equiv&i\,(2\pi)^{4} \int {\rm d}\tilde{q}\,\delta(k+p+q)\,
(k-p)_{\mu}\,\left[\vee^{(1)\,\mu}\right]_{q}\,,\nonumber\\
 A_{\hi kp}^{(2)}&\equiv& (2\pi)^{4}\int {\rm d}\tilde{q}\, {\rm d}\tilde{r}\,\delta(k+p+q+r)\,
\left[\vee^{(2)} \right]_{q,r}\,,\nonumber\\
{\cal B}^T_k &\equiv& (2\pi)^{4}\int {\rm d}\tilde{q}\, {\rm d}\tilde{p}\,
\delta(k+p+q)\,{\cal B}^T_{q,p}\,.
\end{eqnarray}

Once the classical action has been written in the proper 
form~(\ref{eq:gammash}),
we proceed with the functional integration to one loop of the heavy Higgs
particles $\hi$. By using the standard path integral techniques we get the
following result for the effective action,
\begin{equation}
\Gamma_{eff}^{\hi} [V] = \Gamma_0[V]+\frac{i}{2} \Tr \log A_{\hi}
-\frac{1}{2} \langle{\cal B}^T\,A_\hi^{-1}\,{\cal B}\rangle\,,
\label{eq:gammaeffhiint}
\end{equation}
where, 
$$
\langle{\cal B}^T\,A_\hi^{-1}\,{\cal B}\rangle \equiv
\int {\rm d}\tilde{k}\, {\rm d}\tilde{p}\,{\cal B}^T_k\, 
A_{\hi kp}^{-1}\,{\cal  B}_p\,.
$$
In~(\ref{eq:gammaeffhiint}) we have introduced the functional trace which for a
generic matrix operator $C^{ij}(k,p)\equiv C^{ij}_{kp}$ is defined 
by~\cite{Dobado:1997up}:
$$
\Tr C\equiv \sum_i \int {\rm d}\tilde{k} C^{ii}_{kk}\,\,.
$$
Next, by expanding the logarithm and the inverse 
operator in~(\ref{eq:gammaeffhiint}),
the effective 
action can be  written as,
\begin{equation}
\begin{array}{l}
\displaystyle 
\Gamma_{eff}^{\hi} [V] = \Gamma_0[V]+\frac{i}{2} \sum_{k=1}^{\infty} \frac{(-1)^{k+1}}{k} \Tr 
[G_{\hi} (A_{\hi}^{(1)} + A_{\hi}^{(2)})]^{k}-
\frac{1}{2} \sum_{k=0}^{\infty} (-1)^{k} \langle{\cal B}^T\,
[G_{\hi} (A_{\hi}^{(1)} + A_{\hi}^{(2)})]^{k}\,G_{\hi}\,{\cal B}\rangle\,, 
\end{array}
\end{equation}
where $G_{\hi}$ is the heavy Higgs propagator matrix, defined as
$G_{\hi}=(A_{\hi}^{(0)})^{-1}$, and is given in momentum space by,
\begin{equation}
\label{eq:propH}
 G_{\hi kp } =
(2\pi)^4 \delta(k+p) (k^{2} - 
M_{\hi}^{2})^{-1} \,\,,
\end{equation}
with
$$
\displaystyle (q^{2} - M_{\hi}^{2})^{-1} = 
\diag(\frac{1}{q^{2} - m_{H^{1}}^{2}},\frac{1}{q^{2} -
  m_{H^{2}}^{2}}, \frac{1}{q^{2} - m_{H^{o}}^{2}},\frac{1}{q^{2}
  - m_{A^{o}}^{2}})\,.
$$

Finally, if we keep just the terms that contribute to the two, three and four
point $V$ Green functions we get,
\begin{eqnarray}
\label{eq:effHiggs}
\displaystyle 
\Gamma_{eff}^{\hi} [V] &=& 
\Gamma_0[V] -\frac{1}{2} \langle{\cal B}^T G_{\hi}{\cal B} \rangle \nonumber\\
&+&\frac{i}{2} \Tr (G_{\hi} A_{\hi}^{(2)}) - 
\frac{i}{4} \Tr (G_{\hi} A_{\hi}^{(1)})^{2} \nonumber\\
&-& 
\frac{i}{2} \Tr (G_{\hi} A_{\hi}^{(1)} G_{\hi} A_{\hi}^{(2)})+
\frac{i}{6} \Tr (G_{\hi} A_{\hi}^{(1)})^{3} \nonumber\\
&- & \frac{i}{4} \Tr (G_{\hi} A_{\hi}^{(2)})^{2}+
\frac{i}{2} \Tr (G_{\hi} A_{\hi}^{(1)}G_{\hi}  A_{\hi}^{(1)}
G_{\hi} A_{\hi}^{(2)})\nonumber\\
&-&
\frac{i}{8} \Tr (G_{\hi} A_{\hi}^{(1)})^{4}
 +O(V^{5})\,.
\end{eqnarray}
The various contributions can be clearly identified from this expression. The
third and fourth terms give the one-loop contributions to 
the two-point functions; the
two next terms to the three-point functions; and the last three 
terms correspond
to the four-point functions. Notice that there is just one contribution from the
Higgs integration at the tree level. This is the second term 
in eq.~(\ref{eq:effHiggs}) and contributes just to the four point functions. 
Note that it is the unique 
sector that generates a contribution to the electroweak gauge boson functions 
at the tree level. As we have seen
in~\cite{Dobado:1997up,Dobado:1999cz} the integration of sfermions and {\it
  inos} in the effective action for electroweak gauge bosons give only
contributions starting from one-loop level. In addition, notice also that the
resulting effective action in eq.~(\ref{eq:effHiggs}) is gauge independent, as
expected. This is due to the fact that we only integrate the physical Higgs
particles whose interactions with the electroweak gauge bosons are gauge
independent. 

Finally, and for the purpose of illustration, we
have shown in Fig.~\ref{fig:uno} the Feynman diagrams corresponding to the
different terms appearing in the above eq.~(\ref{eq:effHiggs}).
\figuno

\section{The n-point functions of electroweak gauge bosons}
The effective action can be written in terms of the n-point Green's functions
in momentum space, generically as:
\begin{eqnarray}
\label{eq:effp} 
\displaystyle
\Gamma_{eff} [V] &=& \sum_{n} \frac{1}{C_{\scriptstyle {V_{1} V_{2} \ldots V_{n}}}}
\int  {\rm d}\tilde k_{1} \ldots  {\rm d}\tilde k_{n} (2\pi)^{4}
\delta({{\Sigma}_{i=1}^{n}k_{i}}) \times\nonumber\\
&& \Gamma_{\mu \, \nu \ldots  \, \rho}^{V_{1} V_{2} \ldots   V_{n}} 
 (k_{1} \, k_{2} \ldots  \,k_{n})\, V_1^{\mu}(-k_{1})\,V_2^{\nu}(-k_{2}) 
 \,\ldots  V_n^{\rho}(-k_{n})\,, 
\end{eqnarray}
where $C_{\scriptstyle {V_{1} V_{2} \ldots  V_{n}}}$ are the proper combinatorial factors accounting for the identical external
field, and we have assumed the convention of incoming momenta $k_i$ for the
external gauge bosons.

In this section we present the exact results to one-loop for the various
contributions to the effective action of eq.~(\ref{eq:effHiggs}) coming from the
2, 3 and 4 point functions and write them in terms of the standard one-loop
integrals of 't~Hooft, Veltman and
Passarino~\cite{tHooft:1979xw,Passarino:1979jh}. We latter analyze the
asymptotic behaviour of the electroweak bosons Green's functions in the limit
of large Higgs masses. The analysis of the one-loop integrals in the large
masses limit have been done by means of the m-Theorem~\cite{Giavarini:1992xz}. 

After working out the functional traces in eq.~(\ref{eq:effHiggs}) and by
computing the corresponding Feynman integrals in dimensional regularization
we get the following contributions, $\Gamma_{eff}^{H}[V]_{[n]}$, from the
$n=2,\,\,3$ and $4$ point functions respectively\footnote {Notice that in
dimensional reduction the results would be the same, since we are not
integrating out gauge bosons. This also applies to the results of our two
previous papers~\cite{Dobado:1997up,Dobado:1999cz}.},
\begin{eqnarray}
\label{eq:effexactH}
{\Gamma_{eff}^{\hi} [V]}_{[2]} &=&-{\pi}^2 \int {\rm d}\tilde{p}
\,{\rm d}\tilde{k}\,\, \delta(p+k)\,\left\{ \sum_{i} {\left[\vee^{(2)}
\right]}^{ii}_{p,k} A_{0}(m_{i})\right.\nonumber\\
&&\left.+\frac{1}{4}\sum_{i\neq j}
{\left[\vee^{(1) \mu}\right]}^{ij}_{p}\,
{\left[\vee^{(1) \nu}\right]}^{ji}_{k}\,
I^{i\,j}_{\mu\nu}(k, m_i, m_j)\right\}\,,\\
{\Gamma_{eff}^{\hi} [V]}_{[3]} &=&-i{\pi}^2 \int {\rm d}\tilde{p}
\,{\rm d}\tilde{k}\,{\rm d}\tilde{r}\,\, \delta(p+k+r)\,\left\{ 
\sum_{i\neq j}{\left[\vee^{(1) \mu}\right]}^{ji}_{p} 
{\left[\vee^{(2)}\right]}^{ij}_{k,r}\frac{1}{2}
T^{j\,i}_{\mu} (p, m_i, m_j)\right.\nonumber\\
&&\left.+\frac{1}{6}\sum_{i\neq j \neq k}
{\left[\vee^{(1) \mu}\right]}^{ij}_{-p}\,
{\left[\vee^{(1) \nu}\right]}^{jk}_{-k}\,
{\left[\vee^{(1) \sigma}\right]}^{ki}_{-r}\,
T^{i\,j\, k}_{\mu \,\nu \,\sigma}(p,k, m_i, m_j, m_k)\right\}\,,\\
{\Gamma_{eff}^{\hi} [V]}_{[4]} &=&-\frac{1}{2}\sum_{i}\int {\rm d}\tilde{p}\,
{\cal B}^{i}_{p}\, \frac{1}{p^2-m_i^2}\,{\cal B}^{i}_{-p}\nonumber\\
&+&{\pi}^2 \int {\rm d}\tilde{p}
\,{\rm d}\tilde{k}\,{\rm d}\tilde{r}\,\, \delta(p+k+r)\,\left\{ 
\sum_{i,j}{\left[\vee^{(2) \mu}\right]}^{ij}_{-p,-k} 
{\left[\vee^{(2)}\right]}^{ji}_{-r,-t}\,
J^{i\,j}_{p+k}(p+k,m_{i}, m_{j})\right.\nonumber\\
&+&\sum_{i,j,k}
{\left[\vee^{(1) \mu}\right]}^{ij}_{-p}\,
{\left[\vee^{(1) \nu}\right]}^{jk}_{-k}\,
{\left[\vee^{(2) \sigma}\right]}^{ki}_{-r,-t}\,
J^{i\,j\, k}_{\mu \,\nu} 
(p,k,m_{i}, m_{i},m_{k})\nonumber\\
&+&\left. \frac{1}{8}\sum_{i,j,k,l}
{\left[\vee^{(1) \mu}\right]}^{ij}_{-p}\,
{\left[\vee^{(1) \nu}\right]}^{jk}_{-k}\,
{\left[\vee^{(1) \sigma}\right]}^{kl}_{-r}\,
{\left[\vee^{(1) \lambda}\right]}^{li}_{-t}\,
J^{i\,j\,k\,l}_{\mu\,\nu\,\sigma \,\lambda}
(p,k,r,m_{i}, m_{j},m_{k},m_{l})
\right\}\,,\nonumber\\
\end{eqnarray} 
In the above expressions the indices $i,j,k,l$ run from 1 to 4 and correspond to
the four entries in the heavy Higgs matrix $H$ of eq.~(\ref{eq:Hpesado}). 
In these formulas and in the following, a proper symmetrization over the indices 
and momenta of the
external identical fields, although not explicitely shown, must be assumed.
The
one loop integrals  
 $T^{j\,i}_{\mu}$, $T^{i\,j\, k}_{\mu \,\nu \,\sigma}$, 
$J^{i\,j}_{p+k}$, $J^{i\,j\, k}_{\mu \,\nu}$ and 
$J^{i\,j\,k\,l}_{\mu\,\nu\,\sigma \,\lambda}$
are defined in terms of the standard
integrals, $A_0$, $B_{0,\,\mu,\,\mu\nu}$, $C_{0,\,\mu,\,\mu\nu,\,\mu\nu\sigma}$ and
$D_{0,\,\mu\nu,\,\mu\nu\sigma,\,\mu\nu\sigma\lambda}$~\cite{tHooft:1979xw,Passarino:1979jh}
in appendix A of our 
previous work~\cite{Dobado:1999cz}. Similarly, the two-point integral 
$I^{i\,j}_{\mu\nu}$ is defined by,
\begin{equation} 
I^{i\,j}_{\mu\nu}(k, m_i, m_j)=\left[4 B_{\mu\nu}+2k_{\nu} B_{\mu}
+2k_{\mu} B_{\nu}
+k_{\mu}k_{\nu}B_{0}\right](k, m_i, m_j).
\label{eq:Int}
\end{equation}
We refer the reader to~\cite{Dobado:1999cz} for
these and more details on the Feynman integrals.

Finally, from the previous expressions in eq.~(\ref{eq:effexactH}) 
and by using the
definition in~(\ref{eq:effp}) we extract,
after a rather tedious computation, the exact results to one loop for the
two-point, $\Gamma_{\mu\nu}^{V_1V_2}$, three-point,
$\Gamma_{\mu\nu\sigma}^{V_1V_2V_3}$, and four-point,
$\Gamma_{\mu\nu\sigma\lambda}^{V_1V_2V_3V_4}$, Green's functions with all the
possible choices for the external legs, $V_i=A,Z,W^\pm$ which are collected 
in appendix A.  We would like to mention that we have performed all the 
one-loop computations of this paper by the standard diagrammatic method as well
and we have got the same results.  

In the following, and in order to get the n-point Green's functions in 
the decoupling
limit, we use the asymptotic results for the standard one-loop integrals
$A_0(m_i)$, $B_{0,\,\mu,\,\mu\nu}(p,m_i,m_j)$, 
$C_{0,\,\mu,\,\mu\nu,\,\mu\nu\sigma}(p,k,m_i,m_j,m_k)$ and
$D_{0,\,\mu,\mu\nu,\,\mu\nu\sigma,\,\mu\nu\sigma\lambda}
(p,k,r,m_i,m_j,m_k,m_l)$
that we have computed in dimensional regularization and by using the 
m-Theorem~\cite{Giavarini:1992xz},
and were presented in (A.12) of Ref.~\cite{Dobado:1999cz}. These expressions are valid if the
masses $m_{i,\,j,\,k,\,l}$ in the propagators of the integrals are
much larger that the external momenta $p,\,k,\,r$ and if the
differences of the squared masses involved in the same integral are
much smaller than their sums. This last condition is
fulfilled in the present case of  the heavy MSSM Higgs sector, 
even after radiative
corrections are included in the Higgs mass predictions. 
In order to illustrate this point we shortly present in the following the
approximate MSSM Higgs mass values in the decoupling limit that include the
leading radiative corrections. But the conclusions hold even when the full
radiative corrections are employed.
To be more precise,
in the MSSM, using $\ma$ and $\tan\beta$  as input parameters, and
including the leading radiative corrections which can be parametrized
in terms of the quantity,
$$\delta \equiv \frac{3\,G_{\scriptstyle F}}{\sqrt{2} \pi^2}
\frac{m_t^4}{\sin^{2} \beta} \log \left(1+
\frac{M_{\scriptstyle \tilde{Q}}^{2}}{m_t^2}\right)\,,$$ 
the Higgs masses approach the following values, in the decoupling
limit, $\ma\gg m_Z$~\cite{Haber:1996fp},
\begin{eqnarray} 
m_{h^{o}} &\longrightarrow& \sqrt{m_Z^2 \cos^2 2 \beta + \delta \sin^2 \beta}
\,\left[\, 1+\frac{\delta\, m_Z^2\, \cos^2 \beta}
{2\,m^{2}_{A^{o}}\,(m_Z^2 \cos^2 2 \beta 
+\delta \sin^2 \beta)}\right.\nonumber\\
&&\left. -\frac{m_Z^2 \sin^2 2 \beta+\delta \cos^2 \beta}
{2\,m^{2}_{A^{o}}}\,\right]\,,\nonumber\\
m_{H^{o}} &\longrightarrow& m_{A^{o}}\left[1
+\frac{m_Z^2 \sin^2 2 \beta+\delta \cos^2 \beta}
{2\,m^{2}_{A^{o}}}\right]\,,\nonumber\\
m_{H^{\pm}}&\longrightarrow& m_{A^{o}}{\left[1+\frac{m_W^2}{m^{2}_{A^{o}}}
\right]}^{1/2}\,,
\end{eqnarray}
and the mixing angle in the Higgs sector, $\alpha$, approaches to,
$$
\alpha\to\beta-\frac{\pi}{2} \Rightarrow s_{\alpha\beta}\to -1\,\,.
$$
We see, from the previous expressions that indeed, in the decoupling
limit, $\mh$ always stays below a maximum value which can grow up to
about $130\,GeV$ depending on the particular value of $\tan\beta$ and
the common squark mass $M_{\tilde{Q}}$.\footnote{Similar conclusions 
are found if the more general hypothesis of non-common squark mass parameter 
is assumed.} 
The other Higgs bosons, $\Hz$,
$\hpm$ and $\Az$ become very heavy and  approximately degenerate in the
decoupling limit, where $\mH\sim\mhp\sim\ma\gg\mzns$. Therefore the
condition that the squared mass differences for the heavy Higgs sector
of the MSSM are always smaller than their sums 
is largely justified in the decoupling limit both to tree level and in the one
loop approximation. Notice however that in the present computation of the
electroweak Green's functions to one loop level, we use the tree level Higgs
masses in the internal propagators. The use of the radiatively corrected Higgs
masses would be effectively a two loop effect.

Finally, by considering $s_{\alpha\beta}\to-1$ and inserting the
asymptotic expressions of the one loop integrals into
eq.~(\ref{eq:effexactH}) and, after some algebra, we get the Green
functions in the decoupling limit that are collected in appendix A.
These asymptotic results can be summarized by the following generic 
expressions~\cite{Dobado:1999cz},
\begin{equation}
\label{eq:notG} 
\Gamma_{\mu\,\nu...\rho}^{V_1\,V_2...V_n}=
\Gamma_{0\,\mu\,\nu...\rho}^{V_1\,V_2...V_n}
+\Delta \Gamma_{\mu\,\nu...\rho}^{V_1\,V_2...V_n}
\end{equation}
where the subscript 0 refers to the tree level functions, and the one-loop
contributions to the two, three and four-point functions behave,
in the decoupling limit, respectively as follows,
\begin{eqnarray} 
\label{eq:generic} 
\Delta {\Gamma}_{\mu \,\nu}^{V_{1} \, V_{2}}& =&  
\left[\,\Sigma_{(0)}^{V_{1} \, V_{2}}+\Sigma_{(1)}^{V_{1} \, V_{2}} k^2\,\right] 
\, g_{\mu\, \nu} + R_{(0)}^{V_{1} \, V_{2}}\, k_{\mu} k_{\nu}
+O\left(\frac{k^2}{\Sigma m^2},\frac{\Delta m^2}{\Sigma m^2}\right)\nonumber\\
\Delta\Gamma_{\mu \,\nu\,\sigma}^{V_1\,V_2\,V_3}&=&F^{V_1\,V_2\,V_3}
L_{\mu \,\nu\,\sigma}
+O\left(\frac{k^2}{\Sigma m^2},\frac{\Delta m^2}{\Sigma m^2}\right)\nonumber\\
\Delta\Gamma_{\mu\,\nu\,\sigma\,\lambda}^{V_1\,V_2\,V_3\,V_4}&=&
G^{V_1\,V_2\,V_3\,V_4}\beta_{\mu\,\nu\,\sigma\,\lambda}
+O\left(\frac{k^2}{\Sigma m^2},\frac{\Delta m^2}{\Sigma m^2}\right)
\end{eqnarray}
where,
\begin{equation}
\Sigma_{(0)}^{V_{1} \, V_{2}}\rightarrow 0 \,\,,\,\, 
R_{(0)}^{V_{1} \, V_{2}}=-\Sigma_{(1)}^{V_{1} \, V_{2}}\,\,,\,\, 
O\left(\frac{k^2}{\Sigma m^2},
\frac{\Delta m^2}{\Sigma m^2}\right)\rightarrow 0, 
\end{equation}
$k$ denotes generically any of the external momenta and $\Sigma m^2$ and 
$\Delta m^2$ refer generically to sums and differences of Higgs squared masses 
respectively. The relevant content are in the functions 
$\Sigma_{(1)}^{V_{1} \, V_{2}}$, 
$F^{V_1\,V_2\,V_3}$ and $G^{V_1\,V_2\,V_3\,V_4}$ which contain  a 
$\Delta_\epsilon$
proportional term, with $\Delta_\epsilon$ being defined in (A.1), and a finite contribution that is a logarithmic function of
the heavy Higgs masses, $m_{H^0}$, $m_{H^+}$ and $m_{A^0}$. These functions are
precisely the only remnant of the heavy Higgs particles and, therefore,
{\it a priori}, they summarize all the  potential non-decoupling effects of these
particles in the low energy electroweak gauge bosons physics.
In the next section we will show, however, that these apparent non-decoupling 
effects are,
  indeed, 
 non-physical since they do not manifest in the electroweak
observables.

\section{Decoupling of the MSSM Higgs particles {\em {\'a} la Appelquist 
Carazzone}}
In the previous section we have presented the asymptotic results of the 
electroweak gauge boson functions coming from the integration at one loop 
of the heavy MSSM Higgs particles.  We have shown that all the potential
non-decoupling effects of these heavy Higgs particles  manifest as divergent
contributions in $D=4$ and some finite contributions logarithmically 
dependent on the heavy Higgs masses. Furthermore, as can be seen in
eq.~(\ref{eq:generic}), these contributions are both proportional to the tree
level functions, so that we expect them to be finally absorbed by some proper 
redefinition of the low energy SM parameters. 

In this section we are going 
to complete the demonstration of decoupling  
of the MSSM Higgs particles {\em {\'a} la Appelquist 
Carazzone} by finding a particular set of counterterms for the SM 
electroweak parameters
which precisely allow to absorb all the mentioned effects. We will also
show  that these explicit counterterms coincide 
with the expressions of the corresponding on-shell SM counterterms in the
decoupling limit. By using the common language in the renormalization 
context, 
it is equivalent to say that the decoupling at the Green
functions (or effective action) level manifests if (and only if) the on-shell 
prescription for the counterterms is fixed. Of course, once the decoupling is
shown at the electroweak gauge boson functions level, the decoupling in the 
observables with external electroweak gauge bosons is 
automatically ensured, and this latter is obviously independent of the 
renormalization prescription.  

Let us start by stating the condition for decoupling in terms of the 
renormalized electroweak gauge boson functions. As usual, these functions 
are obtained as follows,   
\begin{equation}
\label{eq:LReno}
\Gamma_{R\,\mu \, \nu ... \, \rho}^{\,V_{1} V_{2} ...  V_{n}}(c_{i\,R})=
\Gamma_{0\,\mu \, \nu ... \, \rho}^{\,V_{1} V_{2} ...  V_{n}}(c_{i\,R})+
\Delta\Gamma_{\mu \, \nu ... \, \rho}^{\,V_{1} V_{2} ...  V_{n}}(c_{i\,R})+
\delta\Gamma_{\mu \, \nu ... \, \rho}^{\,V_{1} V_{2} ...  V_{n}}(c_{i\,R})\,,
\end{equation}
where, once more, $\Gamma_{0}$ denote the tree level 
contributions,  $\Delta\Gamma$ are the one-loop contributions, 
and $\delta\Gamma$ represent the contributions from the counterterms of the
SM parameters and wave functions.
All these contributions must be written in terms of the
renormalized parameters that we have denoted here generically by $c_{i\,R}$.
Now, the decoupling of heavy particles {\em {\'a} la Appelquist 
Carazzone} is equivalent to the statement that the renormalized Green functions
are equal to the corresponding tree level functions, evaluated  at the
renormalized parameters, plus corrections that go as inverse powers of the heavy
masses and vanish in the asymptotic limit. Therefore, it implies the following 
conditions,
\begin{equation}
\label{eq:CondAC}
\Delta\Gamma_{\mu \, \nu ... \, \rho}^{\,V_{1} V_{2} ...  V_{n}}(c_{i\,R})+
\delta\Gamma_{\mu \, \nu ... \, \rho}^{\,V_{1} V_{2} ...  V_{n}}(c_{i\,R})
\approx 0\,;\,k^2 \ll m_i^2, \forall i \,, 
\end{equation}
where, for the present case, 
$m_i$ are the heavy Higgs masses, $k$ any of the external momenta, and,   
by $\approx 0$ we mean quantities vanishing in the decoupling limit which 
have been written generically along this paper as being of 
$O\left(\frac{k^2}{\Sigma m^2},\frac{\Delta m^2}{\Sigma m^2}\right)$. 

In order to find the wanted explicit SM counterterms we need to include in 
eq.~(\ref{eq:CondAC}) the asymptotic results presented in the previous section 
for $\Delta \Gamma$, write $\delta \Gamma$ in terms of the SM counterterms and
finally solve the complete system of equations with all the two, three and four 
point functions included.

By using the standard multiplicative renormalization 
procedure~\cite{Bohm:1986rj,Hollik:1990ii}, the 
bare SM electroweak fields and parameters, denoted here by a superscript $0$, 
and the renormalized ones are related by,
\begin{eqnarray} 
\label{eq:renor1} 
\vec{W}^0_{\mu} \equiv Z_W^{1/2}\,\vec{W}_{\mu}\,&\,,&\,
B^0_{\mu} \equiv Z_B^{1/2}\,B_{\mu}\,\,,\,\,
\Phi^0={(Z_{\Phi})}^{\frac{1}{2}}\Phi,\nonumber\\
\xi_W^0 \equiv \xi_W (1+\delta \xi_W)\,&\,,&\,\,
\xi_B^0 \equiv \xi_B (1+\delta \xi_B),\nonumber\\
\label{eq:Zs} 
g^0 \equiv Z_W^{-1/2} \,(g -\delta g)\,&\,,&\,
{g'}^0 \equiv Z_B^{-1/2} \,(g' -\delta g')\,,\,\,\nonumber\\
v^0 = {(Z_{\Phi})}^{\frac{1}{2}}(v-\delta v)\,&\,,&\,\,
Z_i \equiv 1+\delta Z_i\,\,,\,{i\equiv \scriptstyle {A,Z,W,B,\Phi}}\,.
\end{eqnarray}
The counterterms for the physical masses and physical fields are related to
the previous ones by,
\begin{eqnarray}
\delta m_W^2&=&m_W^2(\delta Z_\Phi-2\frac{\delta g}{g}
-2\frac{\delta v}{v}-\delta Z_W)\nonumber\\ 
\delta m_Z^2&=&m_Z^2(\delta Z_\Phi-2\cw^2\frac{\delta g}{g}
-2\sw^2\frac{\delta g'}{g'}
-2\frac{\delta v}{v}-\delta Z_Z)\nonumber\\ 
\delta Z_A &=& \sw^2 \delta Z_W + \cw^2 \delta Z_B\nonumber\\ 
\delta Z_Z &=& \cw^2 \delta Z_W + \sw^2 \delta Z_B\,, 
\label{eq:Zphysical}
\end{eqnarray}
where, as usual, $s_W^2=1-m_W^2/m_Z^2$ and $e=gs_W$.

 The contributions from the various renormalization constants  
 to the two, three and four 
 point functions can be written as~\cite{Bohm:1986rj}, 
\begin{eqnarray} 
\delta \Gamma_{\mu \nu}^{AA}&=& \left[- 
\left(\sw^2 \,\delta Z_W+ \cw^2 \,\delta Z_B\right)\, k^{2}
\right]\,g_{\mu\nu}\nonumber\\ 
&+& \left[\sw^2\left(\frac{\delta\xi_W}{\xi_W}+
\left(1-\frac{1}{\xi_W} \right)\delta Z_W\right)+
\cw^2\left(\frac{\delta\xi_B}{\xi_B}+
\left(1-\frac{1}{\xi_B} \right)\delta Z_B\right)\right]
\,k_{\mu}\,k_{\nu}\,,\nonumber\\
\delta \Gamma_{\mu \nu}^{AZ} &=&  \left[
\frac{\sw}{\cw} m_W^{2} \left(\frac{\delta g'}{g'}-\frac{\delta
    g}{g}\right) -\sw \cw \left(\,\delta Z_W-\delta Z_B\,\right)
  k^{2}\right]\,g_{\mu\nu}\nonumber\\ 
&+& \sw \cw \left[\frac{\delta\xi_W}{\xi_W}+
\left(1-\frac{1}{\xi_W} \right)\delta Z_W-
\frac{\delta\xi_B}{\xi_B}-
\left(1-\frac{1}{\xi_B} \right)
\delta Z_B\right]\,k_{\mu}\,k_{\nu}\,,\nonumber\\
\delta \Gamma_{\mu \nu}^{ZZ}&=& \left[\,\delta m_Z^{2}+
\left(m_Z^{2}-k^{2} \right)\left(\cw^2 \delta Z_W + \sw^2 \delta Z_B
\right)\right]\,g_{\mu\nu}\nonumber\\ 
&+& \left[\cw^2\left(\frac{\delta\xi_W}{\xi_W}+
\left(1-\frac{1}{\xi_W} \right)\delta Z_W\right)+
\sw^2\left(\frac{\delta\xi_B}{\xi_B}+
\left(1-\frac{1}{\xi_B} \right)\delta Z_B\right)\right]
\,k_{\mu}\,k_{\nu}\,,\nonumber\\
\delta \Gamma_{\mu \nu}^{WW}&=& \left[\,\delta m_W^{2}+
\left(m_W^{2}-k^{2} \right)\, \delta Z_W\right]\,g_{\mu\nu}
+ \left[\frac{\delta\xi_W}{\xi_W}+
\left( 1-\frac{1}{\xi_W} \right)\delta Z_W\right]
\,k_{\mu}\,k_{\nu}\,,\nonumber\\
\nonumber\\
\delta \Gamma_{\mu \nu\sigma}^{AW^{+}W^{-}}&=& g\,\sw\,  
{\myL}_{\mu\,\nu\,\sigma} \left[\delta Z_W- 
\frac{\delta g}{g}\,\right]\,\,\,,\,\,\,
\delta \Gamma_{\mu \nu\sigma}^{ZW^{+}W^{-}}= g\,\cw\,  
{\myL}_{\mu\,\nu\,\sigma} \left[\delta Z_W- 
\frac{\delta g}{g}\,\right]\,,\nonumber\\ 
\nonumber\\ 
\delta \Gamma_{\mu\, \nu\,\sigma\,\lambda}^{AAW^{+}W^{-}}&=& -g^2\,\sw^2 
\,{\myss}_{\mu \nu \sigma \lambda} \left[\delta Z_W- 
2\,\frac{\delta g}{g}\,\right]\,,\,\,
\delta \Gamma_{\mu\, \nu\,\sigma\,\lambda}^{AZW^{+}W^{-}}= -g^2\,\sw\,\cw 
\,{\myss}_{\mu \nu \sigma \lambda} \left[\delta Z_W- 
2\,\frac{\delta g}{g}\,\right],\nonumber\\ 
\delta \Gamma_{\mu\, \nu\,\sigma\,\lambda}^{ZZW^{+}W^{-}}&=& -g^2\,\cw^2 
\,{\myss}_{\mu \nu \sigma \lambda} \left[\delta Z_W- 
2\,\frac{\delta g}{g}\,\right]\,,\,\,
\delta \Gamma_{\mu\, \nu\,\sigma\,\lambda}^{W^{+}W^{-}W^{+}W^{-}}= g^2\, 
\,{\myss}_{\mu \sigma \nu \lambda}\, \left[\,\delta Z_W- 
2\,\frac{\delta g}{g}\,\right].\nonumber\\ 
\label{eq:COUNT}
\end{eqnarray} 
 The results for the one-loop contributions to the electroweak gauge
boson functions, presented in the previous section and in appendix A, 
can be rewritten in a more 
simplified form and in terms of just the heavy $m_{A^0}$ mass as follows,
\begin{eqnarray} 
\Delta {\Gamma}_{\mu \,\nu}^{A \, A} =  
  -\frac{e^2}{8\pi^2} \,{\cal K}_{\mu\, \nu}
      \,\Psi_H \,&,&\,
 \Delta {\Gamma}_{\mu \,\nu}^{A \, Z} =  
 \frac{eg}{16\pi^2} \frac{(2s_W^2-1)}{c_W}\, 
  {\cal K}_{\mu\, \nu}\Psi_H ,\nonumber\\
  \Delta {\Gamma}_{\mu \,\nu}^{W \, W} =  
-\frac{g^2}{16\pi^2}\,{\cal K}_{\mu\, \nu}\Psi_H \,&,&\, 
\Delta {\Gamma}_{\mu \,\nu}^{Z \, Z} =  
-\frac{g^2}{16\pi^2} \frac{(2s_W^2-1)^2+1}{2c_W^2}\, 
  {\cal K}_{\mu\, \nu}\Psi_H  ,\nonumber\\ 
 \displaystyle {\Delta {\Gamma}_{\mu \,\nu \,\sigma}^{A W^+ W^-}}= 
\frac{eg^{2}}{16 \pi^{2}}\,{\myL}_{\mu\,\nu\,\sigma}\,
\Psi_H&\,,&
 {\Delta {\Gamma}_{\mu \,\nu \,\sigma}^{Z W^+ W^-}}= 
\frac{g^{3}}{16 \pi^{2}}\,\cw\,{\myL}_{\mu\,\nu\,\sigma}\,
\Psi_H\,,
\nonumber\\
\displaystyle {\Delta {\Gamma}_{\mu \,\nu \,\sigma\,\lambda}^{A A W^+ W^-}}= 
-\frac{e^{2}g^{2}}{16 \pi^{2}}\,{\myss}_{\mu \nu \sigma \lambda}\,
\Psi_H&\,,&
 {\Delta {\Gamma}_{\mu \,\nu \,\sigma\,\lambda}^{A Z W^+ W^-}}= 
-\frac{eg^{3}}{16 \pi^{2}}\,\cw\,{\myss}_{\mu \nu \sigma \lambda}\,
\Psi_H\,,\nonumber\\
\displaystyle {\Delta {\Gamma}_{\mu \,\nu \,\sigma\,\lambda}^{Z Z W^+ W^-}}= 
-\frac{g^{4}}{16 \pi^{2}}\,\cw^{2}\,{\myss}_{\mu \nu \sigma \lambda}\,
\Psi_H&\,,&
\displaystyle {\Delta {\Gamma}_{\mu \,\nu \,\sigma\,\lambda}^{W^+W^- W^+ W^-}}= 
\frac{g^{4}}{16 \pi^{2}}\,{\myss}_{\mu \sigma \nu \lambda}\,
\Psi_H
\label{eq:1loop}
\end{eqnarray}
where,
\begin{eqnarray}
\label{eq:Pito}
\Psi_H \equiv
 \frac{1}{6}\,\left(\,\Delta_{\epsilon} 
-\log \frac{m_{A^0}^2}{\mu_{o}^{2}}\right)\,&,&\,
{\cal K}_{\mu\, \nu} \equiv
 k^2 g_{\mu\, \nu}- 
   k_{\mu} k_{\nu}  
\end{eqnarray}
By plugging the previous results of eqs.~(\ref{eq:Zphysical}) 
through~(\ref{eq:Pito}) into eq.~(\ref{eq:CondAC}) and by solving the system 
we finally find
the following solution for the SM counterterms:\footnote {Similar results have
been found for sfermions, charginos and neutralinos in~\cite{YO:2000}.}
\begin{eqnarray} 
\displaystyle \delta Z_A &=& -
\frac{e^{2}}{8 \pi^{2}} \Psi_H 
\nonumber\\  
\displaystyle \delta m_W^{2}&=& -m_W^{2}\, \delta Z_W  
 = \frac{g^{2}}{16 \pi^{2}} m_W^{2} \Psi_H 
\nonumber\\ 
\displaystyle \delta m_Z^{2}&=& -m_Z^{2}\, \delta Z_Z  
=\frac{g^{2}}{16 \pi^{2}} \frac{m_Z^{2}}{\cw^2}(1-2\sw^2+2\sw^4)\Psi_H,
\label{eq:solcount} 
\end{eqnarray}
and,
\begin{eqnarray} 
\label{eq:other}
\delta \xi_W =\delta Z_W  \, &,&\,
\delta \xi_B =\delta Z_B  \,, \nonumber\\ 
\frac{\delta g'}{g'}\approx 0\,&,&\,\frac{\delta g}{g}\approx 0\,.
\end{eqnarray}
Notice that, as in our previous formulas, 
the results for all the counterterms above have
corrections, not explicitely shown, that vanish in the asymptotic limit of
infinitely heavy $m_{A^0}$.  
   
To finish this section we find interesting to compare the previous results for
the SM counterterms with the corresponding counterterms of the  
on-shell renormalization prescription which are defined, as usual, 
by~\cite{Denner:1993kt}: 
\begin{eqnarray}
\delta m_W^2=-{\rm Re}\,\Sigma^{WW}_T(m_W^2)&\,,\,&
\delta Z_W={\rm Re}\,\frac{\partial\Sigma^{WW}_T(k^2)}
{\partial k^2}|_{k^2=m_W^2}\nonumber
\\
\delta m_Z^2=-{\rm Re}\,\Sigma^{ZZ}_T(m_Z^2)&\,,\,&
\delta Z_Z={\rm Re}\,\frac{\partial\Sigma^{ZZ}_T(k^2)}
{\partial k^2}|_{k^2=m_Z^2}\nonumber
\\
 \delta Z_A={\rm Re}\,\frac{\partial\Sigma^{AA}_T(k^2)}
{\partial k^2}|_{k^2=0}&,&  
\frac{\delta g}{g}=\frac{1}{c_Ws_W}\frac{\Sigma ^{AZ}_T(0)}{m_Z}
\label{eq:onshell}
\end{eqnarray}
plus the solution for  $\delta g'$ that is a consequence of
the $U(1)_Y$ Ward identity,
\begin{equation}
\frac{\delta g'}{g'}=0,
\end{equation}
Notice that, after plugging our asymptotic expressions for 
the $\Sigma^{V_1\,V_2}_T$ functions of appendix A into eq.~(\ref{eq:onshell}),
the solutions for the on-shell counterterms coincide with our solutions 
of eqs.~(\ref{eq:solcount}) and~(\ref{eq:other})
in the decoupling limit.

In summary, we have shown in this section that the heavy MSSM Higgs particles 
decouple from the low energy electroweak gauge boson physics. We have found
as well
that the SM counterterms that are needed to absorb all the (non-physical) 
heavy Higgs effects are precisely the on-shell counterterms, being these
consistently evaluated 
in the decoupling limit. 
\section{Comparison with the SM Higgs boson case}
We present in this section the paradigmatic case of the 
SM heavy Higgs boson and its comparison with the present case of a MSSM 
heavy Higgs sector. It is very well known that the SM Higgs particle
does not decouple from the low energy electroweak physics. The  
logarithmic
dependent terms on the heavy Higgs mass that appear in various 
electroweak precision observables to one-loop, as for instance, $\Delta \rho$, 
$\Delta r$..,  are clear remnants of the non-decoupling 
SM Higgs effects. Indeed, it is precisely this non-decoupling phenomenon that
is after all being responsible for the present upper Higgs mass limit, 
$m_H<230\,GeV$ at $95\% CL$, which is imposed by the present 
data not allowing easily to accommodate a heavy Higgs. 

We present in the following the results of integrating out the heavy SM Higgs 
particle at the one-loop level for the electroweak gauge boson part of the SM
effective action. The corresponding results for the so-called 
effective Electroweak Chiral Lagrangian and the chiral parameters  
were found some years ago 
in~\cite{Herrero:1994nc,Herrero:1995iu,
Espriu:1995rm,Dittmaier:1995cr,Dittmaier:1996ee}. 
We will work here instead   
in the different context of the effective SM action and the 
Appelquist Carazzone Theorem that we have chosen in this paper.

By integrating out the physical Higgs boson particle at the one-loop level 
in the SM,  
and by following the same procedure as outlined in the previous sections, 
we have found the following asymptotic 
results for the two, three and four-point 
electroweak gauge functions, to be valid in the very large Higgs mass limit, 
$M_{H_{SM}} \gg M_Z, \, k$, 
\begin{eqnarray}
\label{eq:SM2p}
&&\Delta \Gamma_{\mu\nu}^{\scriptstyle {AA}} \approx 0\,\,\,\,,\,\,\,
\Delta \Gamma_{\mu\nu}^{\scriptstyle {AZ}} \approx 0\nonumber\\
&&\Delta \Gamma_{\mu\nu}^{\scriptstyle {ZZ}} = \frac{g^{2}}{16\pi^{2}} \,
\frac{1}{2\,\cw^{2}}\,\mhSM^{2}\, \left(\Delta_{\epsilon}-\log  
\frac{\mhSM^{2}}{\mu_{o}^{2}}+1\right)\,g_{\mu\nu}\,,\nonumber\\
&&\Delta \Gamma_{\mu\nu}^{\scriptstyle {WW}} = \frac{g^{2}}{16\pi^{2}} \,
\frac{1}{2}\,\mhSM^{2}\, \left(\Delta_{\epsilon}-\log  
\frac{\mhSM^{2}}{\mu_{o}^{2}}+1\right)\,g_{\mu\nu}\,,\nonumber\\
\nonumber\\
\label{eq:SM3p}
&&\Delta \Gamma_{\mu\,\nu\,\sigma}^{\scriptstyle {AWW}} \approx0\,\,\,\,,\,\,\,
\Delta \Gamma_{\mu\,\nu\,\sigma}^{\scriptstyle {ZWW}} \approx 0\nonumber\\
\nonumber\\
\label{eq:SM4p}
&&\Delta \Gamma_{\mu\,\nu\,\sigma\,\lambda}^{\scriptstyle {AAWW}} \approx 0\,\,\,\,,\,\,\,
\Delta \Gamma_{\mu\,\nu\,\sigma\,\lambda}^{\scriptstyle {AZWW}} \approx 0\nonumber\\
&&\Delta \Gamma_{\mu\,\nu\,\sigma\,\lambda}^{\scriptstyle {ZZWW}}= \frac{g^{4}}{16\pi^{2}} \,
\frac{1}{2\,\cw^{2}}\,\left(\Delta_{\epsilon}-\log  
\frac{\mhSM^{2}}{\mu_{o}^{2}}\right)\,g_{\mu\nu}\,g_{\sigma\lambda}\,,\nonumber\\
&&\Delta \Gamma_{\mu\,\nu\,\sigma\,\lambda}^{\scriptstyle {WWWW}}= \frac{g^{4}}{16\pi^{2}} \,
\frac{1}{2}\, \left(\Delta_{\epsilon}-\log  
\frac{\mhSM^{2}}{\mu_{o}^{2}}\right)\,(g_{\mu\nu}\,g_{\sigma\lambda}+
g_{\mu\lambda}\,g_{\nu\sigma})\,,\nonumber\\
&&\Delta \Gamma_{\mu\nu\sigma\lambda}^{\scriptstyle {ZZZZ}}= \frac{g^{4}}{16\pi^{2}} 
\frac{1}{2\,\cw^{4}}\, \left(\Delta_{\epsilon}-\log  
\frac{\mhSM^{2}}{\mu_{o}^{2}}\right)(g_{\mu\nu}\,g_{\sigma\lambda}+
g_{\mu\lambda}\,g_{\nu\sigma}+g_{\mu\sigma}\,g_{\nu\lambda})\,,
\end{eqnarray}
where $\approx 0$ here means quantities that go with inverse powers 
of the SM Higgs mass and vanish in the asymptotic limit.
 
Next, it is inmediate to find out the corresponding SM counterterms given by,
\begin{eqnarray}
\frac{\delta m_W^2}{m_W^2}&=&\frac{\delta m_Z^2}{m_Z^2}\,=\,
-\frac{g^2}{16\pi^2}\frac{1}{2}\,
\frac{\mhSM^2}{m_W^2}\left(\Delta_{\epsilon}-\log  
\frac{\mhSM^{2}}{\mu_{o}^{2}}+1\right)\,,\nonumber \\
\frac{\delta g}{g}&\approx& 0\,\,,\,\,\frac{\delta g'}{g'}\,\approx \,0\,, \nonumber \\
\delta Z_W &=& \delta\xi_W \,\approx \,0\,\,,\,\, 
\delta Z_B \,=\,\delta\xi_B \,\approx\, 0\,. 
\label{eq:SMcount}
\end{eqnarray} 
By comparing eq.~(\ref{eq:SMcount}) and
eqs.~(\ref{eq:solcount}), (\ref{eq:other}) we already see some  
differences. While in the MSSM all the Higgs mass dependence, in the decoupling 
limit, is logarithmic, in the SM case the dominant contribution to the two point
functions goes with the square of the Higgs mass. 
Another relevant difference is in the four point functions. The results in
eq.~(\ref{eq:SMcount}) show that the one-loop corrections from the SM Higgs 
integration are not proportional to the tree level tensor, 
${\myss}_{\mu \nu \sigma \lambda}$, and, as a consequence, these can not be absorbed by the
SM counterterms. This is a clear indication of the non-decoupling of the Higgs
particle. 

Finally, by substituting the previous results of eqs.~(\ref{eq:SM4p}) and
(\ref{eq:SMcount}) into eq.~(\ref{eq:LReno}) 
, we see that the resulting renormalized SM Green functions at low energies, 
$k \ll \mhSM$, are not all 
equal to the tree level ones evaluated at the renormalized
parameters, as in the MSSM case, but there are some extra terms in the four     
functions given generically by,
\begin{equation}
\label{eq:LRenoSM}
\Gamma_{R\,\mu \, \nu \,\sigma \,\lambda}^{\,V_{1} V_{2} V_{3} V_{4}}(c_{i\,R})-
\Gamma_{0\,\mu \, \nu \,\sigma \,\lambda}^{\,V_{1} V_{2} V_{3} V_{4}}(c_{i\,R})=
a_5 \left( \frac{g^2}{2}W_\mu W^\mu+\frac{g^2}{4c_W^2}Z_\mu Z^\mu \right )^2,
\end{equation}
with,
\begin{equation}
\label{eq:a5}
a_5=\frac{v^2}{8\mhSM^2}+\frac{1}{16\pi^2}\frac{1}{4}\left(\Delta_{\epsilon}-\log  
\frac{\mhSM^{2}}{\mu_{o}^{2}}\right)\,.
\end{equation}
Notice that the value of this effective parameter does not coincide with the
so-called electroweak chiral parameter $a_5$ computed 
in~\cite{Herrero:1994nc,Herrero:1995iu,Espriu:1992vm}. The reason is 
because this
later contains the quantum effects of mixed diagrams with both gauge bosons
and the Higgs particle in the loops which are relevant for the computation of 
the non-decoupling contributions to observables as for
instance $\Delta \rho$. In contrast the result presented in eq.~(\ref{eq:a5})
does not include these mixed diagrams.
 
In summary, the previous eq.~(\ref{eq:LRenoSM}) shows explicitely 
that the decoupling theorem of Appelquist and Carazzone does not apply in the
case of the SM with a very heavy Higgs particle. 

\section{Conclusions}
We have shown in this work that the heavy Higgs Sector of the MSSM composed of
the $H^{\pm}$, $H^0$ and $A^0$ scalar particles decouple from the electroweak
SM gauge boson physics at the one loop level and under the hypothesis that the 
Higgs masses are well above the electroweak gauge boson masses. 
The demonstration has consisted of the computation of the effective action 
for the electroweak gauge bosons that results after the integration to one loop
of the $H^{\pm}$, $H^0$ and $A^0$ Higgs bosons. We have found that, in the 
limit of very large $m_A^0$ as compared to the electroweak scale, all these
one-loop effects can be absorbed into redefinitions of the SM parameters,
more specifically by the counterterms of
eqs.~(\ref{eq:solcount}) and (\ref{eq:other}).

In this decoupling limit the only remnant to low energies is, therefore, 
the light
MSSM Higgs particle $h_0$ with a mass below approximately $130\, GeV$. 
However, 
it is still an open question if all the interactions of 
this light Higgs particle  with 
all the
SM particles, fermions and gauge bosons, in the decoupling limit 
and to all orders in perturbation
theory, are exactly the same as the SM
Higgs particle interactions. In our opinion, it is an interesting subject
that is worth to investigate.   
\section*{Acknowledgments.}
We wish to thank H.E.Haber and W.Hollik for interesting discussions.
S.P. would like to thank J.Guasch for many helpful discussions. 
This work has 
been partially supported by the Spanish Mi\-nisterio de Educaci{\'o}n y 
Cultura 
under projects CICYT AEN97-1678, AEN93-0776 and PB98-0782, and the fellowship AP95 00503301.

\section*{Appendix A.}
\vspace{0.4cm}
\setcounter{equation}{0}
\renewcommand{\theequation}{A.\arabic{equation}}

We present here the exact results for the $2$, $3$ and $4$-point Green 
functions of the
electroweak gauge bosons, and their asymptotic results in the 
decoupling limit, $m_{A^{o}}\gg m_Z$, and with
all the heavy Higgs masses much larger than any of the external momenta.

In order to present these results for the corresponding Green functions,
we use the notation introduced in eq.~(\ref{eq:notG}).  
For brevity, we have omitted the arguments of the 
one-loop integrals and we use the following compact 
notation:
$$I^{1\,4}_{\mu\nu}\equiv I^{1\,4}_{\mu\nu}(k,m_{H_1},m_{A^{o}})\,\,, 
\,\,\,\,I^{3\,2}_{\mu\nu}\equiv I^{3\,2}_{\mu\nu}(k,m_{H^{o}},m_{H_2})$$ 
$$T_{\mu}^{14}\equiv  
T_{\mu}^{14}(p, m_{H^{1}}, m_{A^{o}})\,\,,\,\,\, 
T_{\nu}^{31} \equiv  
T_{\nu}^{31}(k, m_{H^{o}}, m_{H^{1}})\,,$$ 
$$T_{\mu \nu \sigma}^{123}\equiv  
T_{\mu \nu \sigma}^{123}(p,k, m_{H^{1}}, m_{H^{2}},m_{H^{o}})\,\,,\,\,\, 
T_{\nu \sigma \mu}^{231} \equiv  
T_{\nu \sigma \mu}^{231}(k,r, m_{H^{2}},m_{H^{o}}, m_{H^{1}})\,,
\,\,{\mbox {etc.}}$$ 

Let us mention that all the asymptotic expressions below 
have corrections that are
suppressed by inverse powers of the heavy masses,
which vanish in the asymptotic large mass limit. They have
been evaluated to one loop in dimensional regularization, with:
\begin{equation} 
\label{eq:Delta}
\hspace*{0.6cm}  
\displaystyle {\Delta}_\epsilon=\frac{2}{\epsilon }-{\gamma }_{\epsilon}  
+\log (4\pi) \hspace*{0.2cm}, \hspace*{0.2cm} \epsilon = 4-D\,, 
\end{equation} 
and $\mu_{o}$ is the scale of dimensional regularization.

\subsection*{Two-Point Functions}

By following the notation given in eq.~(\ref{eq:notG}) for the $2$-point Green
functions, 
${\Gamma_{0}}_{\,\mu \,\nu}^{V_{1} V_{2}}$ represent the tree 
level contributions which are
written in a covariant arbitrary gauge $R_{\xi}$ as, 
\begin{eqnarray} 
\label{eq:efftree} 
{\Gamma_0}^{V\, V}_{\mu\, \nu} (k) &=& (m_{\scriptscriptstyle V}^{2}-k^{2}) 
 g_{\mu\, \nu} + \left(1 - \frac{1}{\xi_{{\scriptscriptstyle V}}}\right) 
 k_{\mu} k_{\nu} \,\,\,(V=Z,W)\,\,,\nonumber\\ 
{\Gamma_0}^{A\, A}_{\mu\, \nu} &=&  
-k^{2}g_{\mu\, \nu} + \left(1 - \frac{1}{\xi_{{\scriptscriptstyle A}}}\right) 
 k_{\mu} k_{\nu} \,\,, \,\,\, 
{\Gamma_{0}}_{\mu \,\nu}^{V_{1} \, V_{2}}=0 \,\,\mbox{ if }\,\, V_{1}\neq V_{2}\,,  
\end{eqnarray}  
and $\Delta {\Gamma}_{\mu \,\nu}^{V_{1} \, V_{2}}$ are the one-loop
contributions defined in terms of the transverse and longitudinal parts,
$\Sigma_{\scriptscriptstyle T}^{V_{1} \, V_{2}}$ and 
$\Sigma_{\scriptscriptstyle L}^{V_{1} \, V_{2}}$, by:
\begin{eqnarray} 
\label{eq:gasi} 
\Delta {\Gamma}_{\mu \,\nu}^{V_{1} \, V_{2}} &=&  
 \Sigma_{\scriptscriptstyle T}^{V_{1} \, V_{2}} (k) \left( g_{\mu\, \nu}- 
\frac{k_{\mu} k_{\nu}}{k^2}\right)+ \Sigma_{\scriptscriptstyle L}^{V_{1} \, V_{2}} (k)\, 
\frac{k_{\mu} k_{\nu}}{k^2}\,. 
\end{eqnarray} 

The exact results for the one-loop contributions to the
two-point Green functions of the electroweak gauge
bosons are:
\begin{eqnarray} 
\label{eq:higgsAA}  
\displaystyle \Delta \Gamma^{A\, A}_{\mu\, \nu}(k) &=& 
-\frac{e^{2}}{16\,\pi^2}\left\{\left[A_{0}(m_{H_1})+A_{0}(m_{H_2})\,\right] 
\,g_{\mu \nu}- 
\frac{1}{2}\,\left[\,I^{1\,2}_{\mu\nu}+ 
I^{2\,1}_{\mu\nu}\,\right]\,\right\}\\ 
\nonumber\\ 
\label{eq:higgsZZ}  
\displaystyle \Delta \Gamma^{Z\, Z}_{\mu\, \nu} (k) &=&  
-\frac{g^{2}}{16\,\pi^2}\frac{1}{4\,c_{{\scriptscriptstyle W}}^{2}}  
\left\{ c_{{2 \scriptscriptstyle W}}^{2} 
[A_{0}(m_{H_1})+A_{0}(m_{H_2})]\,g_{\mu \nu}+ 
[A_{0}(m_{H^{o}})+A_{0}(m_{A^{o}})]\,g_{\mu \nu}\right.\nonumber\\ 
&&-\frac{1}{2}\, {(2s_{{\scriptscriptstyle W}}^{2}-1)}^{2} 
\,\left[\,I^{1\,2}_{\mu\nu}+I^{2\,1}_{\mu\nu}\,\right] 
- \left.\frac{1}{2}\,s_{\alpha \beta}^{2}\, 
\left[\,I^{3\,4}_{\mu\nu}+I^{4\,3}_{\mu\nu}\,\right]\,\right\}\\ 
\nonumber\\ 
\label{eq:higgsAZ}  
\displaystyle \Delta \Gamma^{A\, Z}_{\mu\, \nu}(k) &=& 
-\frac{e\,g}{16\,\pi^2} \frac{1}{2\,c_{{\scriptscriptstyle W}}}  
\left\{ c_{{2 \scriptscriptstyle W}} 
[A_{0}(m_{H_1})+A_{0}(m_{H_2})]\,g_{\mu \nu} 
+\frac{1}{2}\,(2s_{{\scriptscriptstyle W}}^{2}-1)  
 \,\left[\,I^{1\,2}_{\mu\nu}+I^{2\,1}_{\mu\nu}\,\right]\right\}\nonumber\\ 
\\ 
\label{eq:higgsWW} 
\displaystyle \Delta \Gamma^{WW}_{\mu\, \nu}(k) &=&  
-\frac{g^{2}}{16\,\pi^2}\frac{1}{4}\, 
\left\{{\phantom{!^{1}_{1}}}
[A_{0}(m_{H_1})+A_{0}(m_{H_2})+A_{0}(m_{H^{o}})+A_{0}(m_{A^{o}})]  
\,g_{\mu \nu}\right.\nonumber\\ 
&&-\left. \frac{1}{4}\,\left[\,I^{1\,4}_{\mu\nu}+ 
I^{2\,4}_{\mu\nu}+I^{4\,1}_{\mu\nu}+ I^{4\,2}_{\mu\nu} 
+s_{\alpha \beta}^{2}\left(\,I^{1\,3}_{\mu\nu}+I^{2\,3}_{\mu\nu}+ 
I^{3\,1}_{\mu\nu}+I^{3\,2}_{\mu\nu}\,\right)\,\right]\right\}\,, 
\end{eqnarray} 
where $I^{i\,j}_{\mu\nu}$ has been defined in eq.~(\ref{eq:Int}) and $A_{0}$ 
is the scalar one-loop integral, 
which  is defined in~\cite{tHooft:1979xw,Passarino:1979jh}. 

By using the
asymptotic results of the one-loop integrals that were presented 
in our previous 
work~\cite{Dobado:1997up,Dobado:1999cz}, we obtain the
following asymptotic results for the one-loop heavy Higgs contributions to the
transverse and longitudinal parts:

$\bullet$ \hspace*{0.5cm} $m^{2}_{H^{\pm}} \gg k^{2}$ : 
\begin{eqnarray} 
\label{eq:SumHAA} 
\displaystyle \Sigma_{{\scriptscriptstyle T}}^{{\scriptscriptstyle AA}} 
(k)_{\hi} &=& - \frac{e^{2}}{16\,\pi^2}\frac{k^{2}}{3}\,  
\left( \Delta_{\epsilon} -\log \frac{m^{2}_{H^{+}}}{\mu_{o}^{2}} \right)\,,\\
\nonumber\\  
\label{eq:SumHAZ} 
\displaystyle \Sigma_{{\scriptscriptstyle T}}^{{\scriptscriptstyle AZ}} 
(k)_{\hi} &=& \frac{e\,g}{16\,\pi^2}\, 
\frac{(2\sw^2-1)}{2\cw}\frac{k^{2}}{3}\,  
\left( \Delta_{\epsilon} -\log \frac{m^{2}_{H^{+}}}{\mu_{o}^{2}} \right)\,,
\end{eqnarray}  
 
$\bullet$ \hspace*{0.5cm} $m^{2}_{H^{\pm}}, m^{2}_{H^{o}},  
m^{2}_{A^{o}} \gg k^{2}\,; |m^{2}_{H^{o}}-m^{2}_{A^{o}}| \ll 
|m^{2}_{H^{o}}+m^{2}_{A^{o}}|$: 
\begin{eqnarray} 
\label{eq:SumHZZ} 
\displaystyle \Sigma_{{\scriptscriptstyle T}}^{{\scriptscriptstyle ZZ}} 
(k)_{\hi} &=& \frac{g^{2}}{16\,\pi^2}\,\frac{1}{4\,\cw^2}\left\{\, 
h(m^{2}_{H^{o}},m^{2}_{A^{o}})\right.\nonumber\\ 
&-&\left. \frac{k^{2}}{3}\,\left[{(2\sw^2-1)}^{2} 
\left( \Delta_{\epsilon} -\log \frac{m^{2}_{H^{+}}}{\mu_{o}^{2}} \right) 
+\left( \Delta_{\epsilon} -\log \frac{m^{2}_{H^{o}}+m^{2}_{A^{o}}} 
{\mu_{o}^{2}}\right)\right]\right\}\,,\nonumber\\   
\end{eqnarray} 
 
$\bullet$ \hspace*{0.15cm} $m^{2}_{H^{\pm}}, m^{2}_{H^{o}},  
m^{2}_{A^{o}} \gg k^{2}\,; |m^{2}_{H^{o}}-m^{2}_{H^{\pm}}| \ll 
|m^{2}_{H^{o}}+m^{2}_{H^{\pm}}|\,; |m^{2}_{A^{o}}-m^{2}_{H^{\pm}}| \ll 
|m^{2}_{A^{o}}+m^{2}_{H^{\pm}}|$: 
\begin{eqnarray} 
\label{eq:SumHWW} 
\displaystyle \Sigma_{{\scriptscriptstyle T}}^{{\scriptscriptstyle WW}} 
(k)_{\hi} &=& \frac{g^{2}}{16\,\pi^2}\,\frac{1}{4} \left\{\,\left[\, 
h(m^{2}_{H^{+}},m^{2}_{H^{o}})+h(m^{2}_{H^{+}},m^{2}_{A^{o}}) 
\right]\right.\nonumber\\ 
&-&\left. \frac{k^{2}}{3}\,\left[ 
\left( \Delta_{\epsilon} -\log \frac{m^{2}_{H^{+}}+m^{2}_{H^{o}}} 
{2\mu_{o}^{2}} \right) 
+\left( \Delta_{\epsilon} -\log \frac{m^{2}_{H^{+}}+m^{2}_{A^{o}}} 
{2\mu_{o}^{2}}\right) 
\right]\right\}\,,  
\end{eqnarray}
where $h({m}_{1}^{2}, {m}_{2}^{2})$ is a function defined as:
\begin{equation} 
\label{eq:h} 
h({m}_{1}^{2}, {m}_{2}^{2}) \equiv {m}_{1}^{2} \log 
\frac{2{m}_{1}^{2}}{{m}_{1}^{2} + {m}_{2}^{2}} + 
{m}_{2}^{2} \log \frac{2{m}_{2}^{2}}{{m}_{1}^{2} + {m}_{2}^{2}}\,, 
\end{equation} 
and whose asymptotic behaviour in the large ${m}_{1}$ and
${m}_{2}$ limit,  with
$|{m}_{1}^{2} - {m}_{2}^{2}| \ll |{m}_{1}^{2} + {m}_{2}^{2}|$ is: 
\begin{eqnarray} 
\label{eq:asinh} 
h({m}_{1}^{2}, {m}_{2}^{2}) &\rightarrow& 
\frac{{m}_{1}^{2} - {m}_{2}^{2}}{2} \left[ 
\frac{({m}_{1}^{2} - {m}_{2}^{2})}{({m}_{1}^{2} + {m}_{2}^{2})} + 
O{\left(\frac{{m}_{1}^{2} - {m}_{2}^{2}}{{m}_{1}^{2} 
+ {m}_{2}^{2}}\right)}^{2} \right]. 
\end{eqnarray} 
The above results can be written, in a generic form, as:
$\Sigma_{{\scriptscriptstyle T}}^{{\scriptscriptstyle V_{1} \, V_{2}}} (k)= 
\Sigma_{{\scriptscriptstyle T}\,(0)}^{{\scriptscriptstyle V_{1} \, V_{2}}}+ 
\Sigma_{{\scriptscriptstyle T}\,(1)}^{{\scriptscriptstyle V_{1} \, 
    V_{2}}}\,k^2$, where
$\Sigma_{{\scriptscriptstyle T}\,(0)}^{{\scriptscriptstyle V_{1} \, V_{2}}}$ 
and 
$\Sigma_{{\scriptscriptstyle T}\,(1)}^{{\scriptscriptstyle V_{1} \, V_{2}}}$ 
are $k$ independent functions. The results 
for the corresponding longitudinal parts can be
summarized in short by:
$$\,\Sigma_{{\scriptscriptstyle L}}^{{\scriptscriptstyle V_{1} \, V_{2}}}(k)= 
\Sigma_{{\scriptscriptstyle T}\,(0)}^{{\scriptscriptstyle V_{1} \, V_{2}}}
\,\,\,\,\forall \,{V_{1} \,V_{2}}\,.$$ 
For example,
\begin{equation} 
\label{eq:longHWW} 
\displaystyle \Sigma_{{\scriptscriptstyle L}}^{{\scriptscriptstyle WW}} 
(k)_{\hi}=\frac{g^{2}}{16\,\pi^2}\,\frac{1}{4}\left\{ 
h(m^{2}_{H^{+}},m^{2}_{H^{o}})+h(m^{2}_{H^{+}},m^{2}_{A^{o}})\right\}. 
\end{equation} 

\subsection*{Three-Point Functions}

Analogously to the previous case, we define the three-point Green functions by
following the notation introduced in eq.~(\ref{eq:notG}), 
with ingoing momenta assignments 
$V_{1}^{\mu}(-p)$, $V_{2}^{\nu}(-k)$ and $V_{3}^{\sigma}(-r)$. 
The tree level
contributions, ${\Gamma_{0}}_{\mu \,\nu \,\sigma}^{V_{1} \, V_{2} \, V_{3}}$,
are given by,
\begin{equation} 
\displaystyle {\Gamma_{0}}_{\mu \,\nu \,\sigma}^{A W^+ W^-}=
e\,{\myL}_{\mu \nu \sigma}\,\,, 
\,\,\, {\Gamma_{0}}_{\mu \,\nu \,\sigma}^{Z W^+ W^-}=
g\cw\,{\myL}_{\mu \nu \sigma}\,,
\end{equation} 
with: 
\begin{equation} 
\label{eq:optree} 
{\myL}_{\mu\,\nu\,\sigma}\equiv\left[(k-p)_{\sigma}g_{\mu\, \nu}+ 
(r-k)_{\mu}g_{\nu\, \sigma}+(p-r)_{\nu}g_{\mu\, \sigma}\right]\,, 
\end{equation} 
and the $A W^+ W^-$ and
$Z W^+ W^-$ exact one-loop contributions are:
\begin{eqnarray} 
\label{eq:AWWexactH} 
{\Delta {\Gamma}_{\mu \,\nu \,\sigma\hspace*{0.4cm}\hi} 
^{A W^+ W^-}}&=& -\frac{eg^{2}}{8} \frac{1}{16\,\pi^{2}}\,\left\{ 
s_{\alpha \beta}^2 \left[\,(T_{\sigma}^{13}-T_{\sigma}^{31})\,g_{\mu \nu}+ 
(T_{\nu}^{31}-T_{\nu}^{13})\,g_{\mu \sigma}\right]\right.\nonumber\\ 
&+&\left[\,(T_{\sigma}^{14}-T_{\sigma}^{41})\,g_{\mu \nu}+ 
(T_{\nu}^{41}-T_{\nu}^{14})\,g_{\mu \sigma}\right]\nonumber\\ 
&-& \frac{1}{3} s_{\alpha \beta}^2  \left[ T_{\nu \sigma \mu}^{231}- 
T_{\sigma \nu \mu}^{231}+T_{\sigma \mu \nu}^{321}- 
T_{\nu \mu \sigma}^{321}+T_{\mu \nu \sigma}^{123}-T_{\mu \sigma \nu}^{123} 
\right]\nonumber\\ 
&-& \left.\frac{1}{3}\left[T_{\nu \sigma\mu}^{142}- 
T_{\sigma \nu \mu}^{142}+T_{\sigma \mu \nu}^{412}- 
T_{\nu \mu \sigma}^{412}+T_{\mu \nu \sigma}^{124}-T_{\mu \sigma \nu}^{124} 
\right]\right\}\,,\\
\nonumber\\  
\label{eq:ZWWexactH} 
\displaystyle {\Delta {\Gamma}_{\mu \,\nu \,\sigma\hspace*{0.4cm}\hi} 
^{Z W^+ W^-}}&=& \frac{g^{3}}{8\,\cw} \frac{1}{16\,\pi^{2}}\,\left\{ 
s_{\alpha \beta}^2 \sw^2 \left[\,(T_{\sigma}^{13}-T_{\sigma}^{31})\,g_{\mu \nu}+ 
(T_{\nu}^{31}-T_{\nu}^{13})\,g_{\mu \sigma}\right]\right.\nonumber\\ 
&+&\sw^2 \left[\,(T_{\sigma}^{14}-T_{\sigma}^{41})\,g_{\mu \nu}+ 
(T_{\nu}^{41}-T_{\nu}^{14})\,g_{\mu \sigma}\right]\nonumber\\ 
&-& \frac{1}{6} s_{\alpha \beta}^2  (2\sw^2-1)\left[ 
T_{\nu \sigma \mu}^{231}- 
T_{\sigma \nu \mu}^{231}+T_{\sigma \mu \nu}^{321}- 
T_{\nu \mu \sigma}^{321}+T_{\mu \nu \sigma}^{123}-T_{\mu \sigma \nu}^{123} 
\right]\nonumber\\ 
&-& \frac{1}{6}(2\sw^2-1)\left[T_{\nu \sigma\mu}^{142}- 
T_{\sigma \nu \mu}^{142}+T_{\sigma \mu \nu}^{412}- 
T_{\nu \mu \sigma}^{412}+T_{\mu \nu \sigma}^{124}-T_{\mu \sigma \nu}^{124} 
\right]\nonumber\\ 
&+& \left.\frac{1}{6} s_{\alpha \beta}^2 \left[ 
T_{\mu \nu \sigma}^{341}-T_{\mu \sigma \nu}^{341}+ 
T_{\mu \nu \sigma}^{431}-T_{\mu \sigma \nu}^{431}+ 
T_{\sigma \mu \nu}^{143}-T_{\nu \mu \sigma}^{143}\right.\right.\nonumber\\ 
&+&\left.\left.T_{\sigma \mu \nu}^{134}-T_{\nu \mu \sigma}^{134}+ 
T_{\nu \sigma \mu}^{413}-T_{\sigma \nu \mu}^{413}+ 
T_{\nu \sigma \mu}^{314}-T_{\sigma \nu \mu}^{314}\,\right]\,\right\}\,,
\end{eqnarray} 
where $T_{\mu}^{ij}$ and $T_{\mu \nu \sigma}^{ijk}$ are the one-loop integrals
as defined in~\cite{Dobado:1999cz}.
 
By using the asymptotic results of the above mentioned integrals,  
we have obtained the following expressions for the
three-point functions in the decoupling limit:
\begin{eqnarray} 
\label{eq:limHaww} 
\displaystyle {\Delta {\Gamma}_{\mu \,\nu \,\sigma\hspace*{0.4cm}\hi} 
^{A W^+ W^-}}&=& \frac{1}{16\pi^2}\,\frac{eg^2}{12}\,{\myL}_{\mu\,\nu\,\sigma}\, 
\left\{\, 
2{\Delta}_\epsilon-\log \frac{2m_{H^{+}}^{2}+m_{H^{o}}^{2}}{3\mu_{o}^{2}} 
-\log \frac{2m_{H^{+}}^{2}+m_{A^{o}}^{2}}{3\mu_{o}^{2}}\, 
\right\}\,,\nonumber\\ 
\\ 
\label{eq:limHzww} 
\displaystyle {\Delta {\Gamma}_{\mu \,\nu \,\sigma\hspace*{0.4cm}\hi} 
^{Z W^+ W^-}} &=&\frac{1}{16\pi^2}\,\frac{g^3}{6\cw}\,{\myL}_{\mu\,\nu\,\sigma}\, 
\left\{ 
\cw^2 {\Delta}_\epsilon-\frac{1}{2}\,\log  
\frac{m_{H^{+}}^{2}+m_{H^{o}}^{2}+m_{A^{o}}^{2}}{3\mu_{o}^{2}}\right.\nonumber\\ 
&&-\left.\frac{c_{\scriptstyle {2W}}}{4}\,\left( 
\log \frac{2m_{H^{+}}^{2}+m_{H^{o}}^{2}}{3\mu_{o}^{2}}+ 
\log \frac{2m_{H^{+}}^{2}+m_{A^{o}}^{2}}{3\mu_{o}^{2}}\right)\right\}\,. 
\end{eqnarray} 

\subsection*{Four-Point Functions}

Finally, for the $4$-point Green functions,
$\Gamma_{\mu \,\nu \,\sigma\,\lambda}^{V_{1} \, V_{2} \, V_{3}\,V_{4}}$,
with ingoing momenta assignments
$V_{1}^{\mu}(-p)$, $V_{2}^{\nu}(-k)$, $V_{3}^{\sigma}(-r)$ and
$V_{4}^{\lambda}(-t)$ we have obtained the results presented below. 
The tree level corresponding contributions different from zero are:
\begin{eqnarray} 
\displaystyle \Gamma_{0\,\,\mu \,\nu \,\sigma\,\lambda}^{AAW^+ W^-}=-e^{2} 
{\myss}_{\mu \nu \sigma \lambda}&,& 
\Gamma_{0\,\,\mu \,\nu \,\sigma\,\lambda}^{AZW^+W^-}= 
-g^{2}\sw\cw{\myss}_{\mu \nu \sigma \lambda}\,, \nonumber\\ 
\nonumber\\ 
\displaystyle \Gamma_{0\,\,\mu \,\nu \,\sigma\,\lambda}^{ZZW^+ W^-}= 
-g^{2}\cw^{2}{\myss}_{\mu \nu \sigma \lambda}&,& 
\Gamma_{0\,\,\mu \,\nu \,\sigma\,\lambda}^{W^+ W^-W^+W^-}= 
g^{2}{\myss}_{\mu \sigma \nu \lambda}\,\,, 
\end{eqnarray} 
where ${\myss}_{\mu \sigma \nu \lambda}$ is defined by,
\begin{equation} 
\label{eq:optree4} 
{\myss}_{\mu\,\nu\,\sigma\,\lambda}\equiv\left[ 2g_{\mu \,\nu}g_{\sigma \,\lambda}- 
g_{\mu \,\sigma}g_{\nu \,\lambda}-g_{\mu \,\lambda}g_{\nu \,\sigma}\right]\,, 
\end{equation} 
and the exact results for the one-loop contributions of the heavy Higgs sector,
$\Delta {{\Gamma}_{\mu \,\nu \,\sigma\,\lambda\hspace*{0.4cm}\hi}^
{V_{1} \, V_{2} \, V_{3}\, V_{4}}}$, are the following:
\begin{eqnarray}
\label{eq:AAWWexactHiggs}
\displaystyle && {\Delta {\Gamma}_{\mu \,\nu \,\sigma\,\lambda
\hspace*{0.4cm}\hi}^{A A W^+ W^-}}=\frac{e^2 g^2}{16 \pi^2}
\bigg\{\, g_{\mu\nu}g_{\sigma\lambda}  J_{p+k}^{11}
+\frac{1}{4}\left(
g_{\mu\sigma}g_{\nu\lambda} J_{p+r}^{14} 
+g_{\nu\sigma}g_{\mu\lambda} J_{k+r}^{14}\right)\nonumber\\&&
+\frac{\sab^2}{4}\left(
g_{\mu\sigma}g_{\nu\lambda} J_{p+r}^{31}
+g_{\nu\sigma}g_{\mu\lambda} J_{k+r}^{31}\right)
-\frac{1}{2} g_{\sigma\lambda} \big[J_{\mu\nu}^{111}+J_{\nu\mu}^{111}\big]\nonumber\\&&
-\frac{\sab^2}{4} 
\big[ g_{\sigma\nu} J_{\mu\lambda}^{113} +g_{\sigma\mu} J_{\nu\lambda}^{113}
     + g_{\lambda\nu} J_{\mu\sigma}^{113} +g_{\lambda\mu} J_{\nu\sigma}^{113}
\big]\nonumber\\&&
-\frac{1}{4} 
\big[  g_{\sigma\nu} J_{\mu\lambda}^{114}   +g_{\sigma\mu} J_{\nu\lambda}^{114}
    + g_{\lambda\nu} J_{\mu\sigma}^{114} +g_{\lambda\mu} J_{\nu\sigma}^{114}
\big]
-\frac{\sab^2}{2} g_{\mu\nu}
J_{\lambda\sigma}^{311}-\frac{1}{2}g_{\mu\nu}J_{\lambda\sigma}^{141}\nonumber\\&&
+\frac{\sab^2}{4} \Jmnls^{1113}
+\frac{1}{4}\, \big[\, \Jmlsn^{1141}+\Jmsln^{1141}\,\big]\bigg\}\,,\\
\nonumber\\
\label{eq:AZWWexactHiggs}
 &&\displaystyle {\Delta {\Gamma}_{\mu \,\nu \,\sigma\,\lambda
\hspace*{0.4cm}\hi}^{A Z W^+ W^-}}=\frac{e g^3}{32 \pi^2\cw}
\bigg\{
\cdw g_{\mu\nu} g_{\sigma\lambda} J_{p+k}^{11} 
-\frac{\sw^2}{2}\left(g_{\mu\sigma}g_{\nu\lambda}J_{p+r}^{14}
-g_{\nu\sigma}g_{\mu\lambda}J_{k+r}^{14}\right)\nonumber\\&&
-\frac{\sab^2\sw^2}{2}\left(g_{\mu\sigma}g_{\nu\lambda}J_{p+r}^{31}
-g_{\nu\sigma}g_{\mu\lambda}J_{k+r}^{31}\right)
-\frac{\cdw}{2}
g_{\sigma\lambda}\big[J_{\mu\nu}^{111}+J_{\nu\mu}^{111}\big]\nonumber\\&&
+\frac{\sab^2\sw^2}{2}\big[\,g_{\nu\lambda}J_{\mu\sigma}^{113}+
g_{\nu\sigma}J_{\mu\lambda}^{113}\,\big]
-\frac{\sab^2\,\cdw}{4}
\big[\,g_{\mu\lambda}J_{\nu\sigma}^{113}+
g_{\mu\sigma}J_{\nu\lambda}^{113}\,\big]\nonumber\\&& 
+\frac{\sw^2}{2}\big[g_{\nu\lambda}J_{\mu\sigma}^{114}+
g_{\nu\sigma}J_{\mu\lambda}^{114}\big]
-\frac{\cdw}{4}
\big[g_{\mu\lambda}J_{\nu\sigma}^{114}+g_{\mu\sigma}J_{\nu\lambda}^{114}\big] 
-\frac{\sab^2\cdw}{2}g_{\mu\nu}J_{\lambda\sigma}^{311}\nonumber\\&&
-\frac{\cdw}{2}g_{\mu\nu}J_{\lambda\sigma}^{141}
+\frac{\sab^2}{4}\big[\,g_{\mu\lambda}J_{\nu\sigma}^{134}+
g_{\mu\sigma}J_{\nu\lambda}^{134}+g_{\mu\lambda}J_{\nu\sigma}^{431}+
g_{\mu\sigma}J_{\nu\lambda}^{431}\,\big]\nonumber\\&&
+\frac{\sab^2\,\cdw}{4}\big[\,\Jmnsl^{1113}+\Jmnls^{1113}\,\big]
+\frac{\cdw}{4}\big[\,\Jmlsn^{1141}+\Jmsln^{1141}\,\big]
-\frac{\sab^2}{4}\big[\,\Jnsml^{1134}+\Jnlms^{1134}\,\big]\bigg\}\,,\\
\nonumber\\
\label{eq:ZZWWexactHiggs}
\displaystyle && {\Delta {\Gamma}_{\mu \,\nu \,\sigma\,\lambda
\hspace*{0.4cm}\hi}^{Z Z W^+ W^-}}=\frac{g^4}{64 \pi^2\cw^2}
\bigg\{ \cdw^2 g_{\mu\nu} g_{\sigma\lambda} J_{p+k}^{11} 
+\frac{1}{2}g_{\mu\nu} g_{\sigma\lambda}
\big[J_{p+k}^{33}+J_{p+k}^{44}\big]\nonumber\\&&
+\sw^4\left( g_{\mu\sigma} g_{\lambda\nu}J_{p+r}^{14}
+g_{\mu\lambda} g_{\nu\sigma} J_{k+r}^{14}\right)
+\sab^2\sw^4\left( g_{\mu\sigma} g_{\lambda\nu} J_{p+r}^{31}
+g_{\mu\lambda} g_{\nu\sigma}J_{k+r}^{31}\right)\nonumber\\&&
-\frac{\cdw^2}{2}g_{\sigma\lambda}\big[J_{\mu\nu}^{111}+J_{\nu\mu}^{111}\big]
-\frac{\sab^2}{2}\, g_{\mu\nu}\, J_{\sigma\lambda}^{133}
-\frac{1}{2}\, g_{\mu\nu}\,J_{\sigma\lambda}^{414}\nonumber\\&&
+\frac{\sab^2\,\sw^2\,\cdw}{2}\big[\,
g_{\mu\lambda}J_{\nu\sigma}^{113}+g_{\nu\lambda}J_{\mu\sigma}^{113}
+g_{\mu\sigma}J_{\nu\lambda}^{113}+g_{\nu\sigma}J_{\mu\lambda}^{113}
\,\big]\nonumber\\&&
+\frac{\sw^2\,\cdw}{4}\big[\,
g_{\mu\lambda}J_{\nu\sigma}^{114}+g_{\nu\lambda}J_{\mu\sigma}^{114}
+g_{\mu\sigma}J_{\nu\lambda}^{114}+g_{\nu\sigma}J_{\mu\lambda}^{114}
\,\big]\nonumber\\&&
-\frac{\sab^2}{2}\left(g_{\sigma\lambda}\,J_{\mu\nu}^{434}
+g_{\sigma\lambda}\,J_{\mu\nu}^{343}\right)
-\frac{\cdw^2}{2}\,g_{\mu\nu}\,J_{\lambda\sigma}^{141}\nonumber\\&&
-\frac{\sab^2\sw^2}{2}\big[
g_{\nu\lambda}J_{\mu\sigma}^{431}+g_{\mu\lambda}J_{\nu\sigma}^{431}
+g_{\nu\sigma}J_{\mu\lambda}^{431}+g_{\mu\sigma}J_{\nu\lambda}^{431}
\nonumber\\&&\hspace{5cm}
+g_{\mu\lambda}J_{\nu\sigma}^{134}+g_{\nu\lambda}J_{\mu\sigma}^{341}
+g_{\mu\sigma}J_{\nu\lambda}^{134}+g_{\nu\sigma}J_{\mu\lambda}^{341}\big]\nonumber\\&&
+\frac{\sab^2\cdw^2}{4}\, \big[\, \Jmnsl^{1113} 
+ \Jmnls^{1113} \,\big]
+\frac{\cdw^2}{4}\, \big[\, \Jmlsn^{1141}+\Jmsln^{1141}\,\big]\nonumber\\&&
+\frac{\sab^2}{4}\,\big[\,\Jmnsl^{4341}+\Jmnls^{4341}\,\big]
+\frac{\sab^4}{4}\,\big[\,\Jmnsl^{3431}+\Jmnls^{3431}\,\big]\nonumber\\&&
-\frac{\sab^2\,\cdw}{4}\,\big[\,
\Jnsml^{4311}+\Jmsnl^{4311}+\Jnlms^{4311}+\Jmlns^{4311}\,\big]\bigg\}\,, 
\end{eqnarray}
\begin{eqnarray}
\label{eq:WWWWexactHiggs}
\displaystyle && {\Delta {\Gamma}_{\mu \,\nu \,\sigma\,\lambda
\hspace*{0.4cm}\hi}^{W^+ W^- W^+ W^-}}=\frac{g^4}{64 \pi^2}
\bigg\{ g_{\mu\nu} g_{\sigma\lambda}\,
\big[\, J_{p+k}^{11}+\frac{1}{2}J_{p+k}^{33}+\frac{1}{2}J_{p+k}^{44}
\,\big]\nonumber\\&&
+ g_{\sigma\nu} g_{\mu\lambda}
\big[\,J_{k+r}^{11}+\frac{1}{2}J_{k+r}^{33}+\frac{1}{2}J_{k+r}^{44}\,\big]
+\frac{1}{2}\big[\,
g_{\nu\sigma}J_{\mu\lambda}^{414}-g_{\nu\mu}J_{\sigma\lambda}^{414}
-g_{\lambda\sigma}J_{\mu\nu}^{414}+g_{\lambda\mu}J_{\sigma\nu}^{414}
\,\big]\nonumber\\&&
-\frac{\sab^2}{2}\big[
g_{\nu\sigma}\left(J_{\mu\lambda}^{313}+J_{\lambda\mu}^{131}\right)
+g_{\nu\mu}\left(J_{\sigma\lambda}^{313}+J_{\lambda\sigma}^{131}\right)
+g_{\lambda\sigma}\left(J_{\mu\nu}^{313}+J_{\nu\mu}^{131}\right)
+g_{\lambda\mu}\left(J_{\sigma\nu}^{313}+J_{\nu\sigma}^{131}\right)
\big]\nonumber\\&&
-\frac{\sab^2}{2}\big[\,
-g_{\nu\sigma}J_{\lambda\mu}^{141}+g_{\nu\mu}J_{\lambda\sigma}^{141}
+g_{\lambda\sigma}J_{\nu\mu}^{141}-g_{\lambda\mu}J_{\nu\sigma}^{141}
\,\big]\nonumber\\&&
+\frac{\sab^4}{4}\big[\,\Jmnsl^{3131}+\Jmlsn^{1313}+
\Jmsnl^{1313}+\Jmsln^{1313}\,\big]
+\frac{1}{4}\big[\,\Jmnsl^{4141}+\Jmlsn^{1414}+
\Jmsnl^{1414}+\Jmsln^{1414}\,\big]\nonumber\\&&
+\frac{\sab^2}{4}\big[
\Jmlsn^{1413}+\Jmlsn^{3141}+\Jmlsn^{1314}+\Jmlsn^{4131}
-\Jmsln^{1314}-\Jmsnl^{1314}-\Jmsln^{1413}-\Jmsnl^{1413}
\,\big]\bigg\}\,.
\end{eqnarray}
Here, $J_{p+k}^{ij}\,,\,J_{\mu\nu}^{ijk}$ and $\Jmnsl^{ijkl}$ are the 
one-loop integrals
given in~\cite{Dobado:1999cz}.

The asymptotic results of the above contributions in the 
decoupling limit can be written as:
\begin{eqnarray} 
\label{eq:AAWWHiggs} 
\displaystyle {\Delta {\Gamma}_{\mu \,\nu \,\sigma\,\lambda\hspace*{0.4cm}\hi} 
^{A A W^+ W^-}}&=& \frac{e^{2}g^{2}}{16 \pi^{2}}\,\bigg\{  
-{\myss}_{\mu \nu \sigma \lambda} \frac{1}{6} \Delta_{\epsilon}+ 
g_{\mu \,\nu}\,g_{\sigma \,\lambda}\, 
g_{1}(m_{H^{+}},m_{H^{o}},m_{A^{o}})\nonumber\\ 
&+& (g_{\mu \,\sigma}\,g_{\nu \,\lambda}+g_{\mu \,\lambda}g_{\nu \,\sigma}) 
\,g_{2}(m_{H^{+}},m_{H^{o}},m_{A^{o}})\bigg\} \\ 
\nonumber\\
\label{eq:AZWWHiggs} 
\displaystyle {\Delta {\Gamma}_{\mu \,\nu \,\sigma\,\lambda\hspace*{0.4cm}\hi} 
^{A Z W^+ W^-}}&=&-\frac{eg^{3}}{16 \pi^{2}}  
\,\frac{1}{2\,\cw}\bigg\{  
\,{\myss}_{\mu \nu \sigma \lambda}\,\frac{\cw^2}{3} \Delta_{\epsilon}+ 
 g_{\mu \,\nu}\,g_{\sigma \,\lambda}\, 
g_{3}(m_{H^{+}},m_{H^{o}},m_{A^{o}})\nonumber\\ 
&+&  (g_{\mu \,\sigma}\,g_{\nu \,\lambda} 
+g_{\mu \,\lambda}g_{\nu \,\sigma}) 
\,g_{4}(m_{H^{+}},m_{H^{o}},m_{A^{o}})\bigg\} \\ 
\nonumber\\ 
\label{eq:ZZWWHiggs} 
\displaystyle {\Delta {\Gamma}_{\mu \,\nu \,\sigma\,\lambda\hspace*{0.4cm}\hi} 
^{Z Z W^+ W^-}}&=&-\frac{g^{4}}{16 \pi^{2}} \,\frac{1}{2\,\cw^2}\,\bigg\{  
\,{\myss}_{\mu \nu \sigma \lambda} \frac{\cw^4}{3} \Delta_{\epsilon}+ 
g_{\mu \,\nu}\,g_{\sigma \,\lambda}\, 
g_{5}(m_{H^{+}},m_{H^{o}},m_{A^{o}})\nonumber\\ 
&+& (g_{\mu \,\sigma}\,g_{\nu \,\lambda}+g_{\mu \,\lambda} 
g_{\nu \,\sigma}) 
\,g_{6}(m_{H^{+}},m_{H^{o}},m_{A^{o}})\bigg\} \\ 
\nonumber\\ 
\label{eq:WWWWHiggs} 
\displaystyle {\Delta {\Gamma}_{\mu \,\nu \,\sigma\,\lambda\hspace*{0.9cm}\hi} 
^{W^+ W^- W^+ W^-}}&=&\frac{g^{4}}{16 \pi^{2}} \frac{1}{4}
\,\bigg\{ \,{\myss}_{\mu \sigma\nu  \lambda}\,\frac{2}{3}\, \Delta_{\epsilon}+ 
g_{\mu \,\sigma}\,g_{\nu \,\lambda}\, 
g_{7}(m_{H^{+}},m_{H^{o}},m_{A^{o}})\nonumber\\ 
&+& (g_{\mu \,\nu}\,g_{\sigma \,\lambda}+g_{\mu \,\lambda} 
g_{\nu \,\sigma}) 
\,g_{8}(m_{H^{+}},m_{H^{o}},m_{A^{o}})\bigg\} 
\end{eqnarray}  
where the $g_{i}(m_{H^{+}},m_{H^{o}},m_{A^{o}})\,\,(i=1\ldots 8)$ functions
are given by,
\begin{eqnarray}
g_{1}&=& \frac{1}{2} \log \frac{2m_{H^{+}}^{2}+m_{A^{o}}^{2}}{3\mu_{o}^{2}}
+\frac{1}{2} \log \frac{2m_{H^{+}}^{2}+m_{H^{o}}^{2}}{3\mu_{o}^{2}}\nonumber\\ 
&-&3\frac{1}{3} \log \frac{3m_{H^{+}}^{2}+m_{A^{o}}^{2}}{4\mu_{o}^{2}}
-\frac{1}{3} \log \frac{3m_{H^{+}}^{2}+m_{H^{o}}^{2}}{4\mu_{o}^{2}}
\,,\nonumber\\
g_{2}&=& -\frac{1}{4} \log \frac{m_{H^{+}}^{2}+m_{A^{o}}^{2}}{2\mu_{o}^{2}}
-\frac{1}{4} \log \frac{m_{H^{+}}^{2}+m_{H^{o}}^{2}}{2\mu_{o}^{2}}
+\frac{1}{2} \log \frac{2m_{H^{+}}^{2}+m_{A^{o}}^{2}}{3\mu_{o}^{2}}\nonumber\\
&+&\frac{1}{2} \log \frac{2m_{H^{+}}^{2}+m_{H^{o}}^{2}}{3\mu_{o}^{2}} 
-\frac{1}{3} \log \frac{3m_{H^{+}}^{2}+m_{A^{o}}^{2}}{4\mu_{o}^{2}}
-\frac{1}{3} \log \frac{3m_{H^{+}}^{2}+m_{H^{o}}^{2}}{4\mu_{o}^{2}}
\,,\nonumber\\ 
g_{3}&=& -\frac{\cdw}{2} \log 
\frac{2m_{H^{+}}^{2}+m_{A^{o}}^{2}}{3\mu_{o}^{2}}
-\frac{\cdw}{2} \log \frac{2m_{H^{+}}^{2}+m_{H^{o}}^{2}}{3\mu_{o}^{2}}
+\frac{\cdw}{3} \log 
\frac{3m_{H^{+}}^{2}+m_{A^{o}}^{2}}{4\mu_{o}^{2}}\nonumber\\
&+&\frac{\cdw}{3} \log \frac{3m_{H^{+}}^{2}+m_{H^{o}}^{2}}{4\mu_{o}^{2}} 
-\frac{1}{3} \log \frac{2m_{H^{+}}^{2}+m_{H^{o}}^{2}+m_{A^{o}}^{2}}
{4\mu_{o}^{2}}\,,\nonumber\\ 
g_{4}&=& -\frac{\sw^2}{2} \log 
\frac{m_{H^{+}}^{2}+m_{A^{o}}^{2}}{2\mu_{o}^{2}}
-\frac{\sw^2}{2} \log 
\frac{m_{H^{+}}^{2}+m_{H^{o}}^{2}}{2\mu_{o}^{2}}
+\left(\sw^2-\frac{1}{4}\right)\log 
\frac{2m_{H^{+}}^{2}+m_{H^{o}}^{2}}{3\mu_{o}^{2}}\nonumber\\
&+&\left(\sw^2-\frac{1}{4}\right)\log 
\frac{2m_{H^{+}}^{2}+m_{A^{o}}^{2}}{3\mu_{o}^{2}}
 +\frac{\cdw}{3} \log 
\frac{3m_{H^{+}}^{2}+m_{H^{o}}^{2}}{4\mu_{o}^{2}}
+\frac{\cdw}{3} \log 
\frac{3m_{H^{+}}^{2}+m_{A^{o}}^{2}}{4\mu_{o}^{2}}\nonumber\\ 
&+& \frac{1}{2} \log \frac{m_{H^{+}}^{2}+m_{H^{o}}^{2}+m_{A^{o}}^{2}}
{3\mu_{o}^{2}}
-\frac{1}{3} \log \frac{2m_{H^{+}}^{2}+m_{H^{o}}^{2}+m_{A^{o}}^{2}}
{4\mu_{o}^{2}}\,,\nonumber\\
g_{5}&=& \frac{1}{4}\log \frac{m_{H^{o}}^{2}}{\mu_{o}^{2}}
+\frac{1}{4}\log \frac{m_{A^{o}}^{2}}{\mu_{o}^{2}}
-\frac{1}{4}\log \frac{2m_{H^{o}}^{2}+m_{H^{+}}^{2}}{3\mu_{o}^{2}}
-\frac{1}{4}\log \frac{2m_{A^{o}}^{2}+m_{H^{+}}^{2}}{3\mu_{o}^{2}}\nonumber\\
&-&\frac{\cdw^2}{4}\log 
\frac{2m_{H^{+}}^{2}+m_{A^{o}}^{2}}{3\mu_{o}^{2}}-
\frac{\cdw^2}{4}\log 
\frac{2m_{H^{+}}^{2}+m_{H^{o}}^{2}}{3\mu_{o}^{2}}-
\frac{1}{4} \log \frac{m_{H^{o}}^{2}+2m_{A^{o}}^{2}}{3\mu_{o}^{2}}\nonumber\\
&-&\frac{1}{4} \log \frac{2m_{H^{o}}^{2}+m_{A^{o}}^{2}}{3\mu_{o}^{2}}
+\frac{\cdw^2}{6} \log 
\frac{3m_{H^{+}}^{2}+m_{H^{o}}^{2}}{4\mu_{o}^{2}}+\frac{\cdw^2}{6} \log 
\frac{3m_{H^{+}}^{2}+m_{A^{o}}^{2}}{4\mu_{o}^{2}}\nonumber\\ 
&+& \frac{1}{6} \log \frac{m_{H^{+}}^{2}+m_{H^{o}}^{2}+2m_{A^{o}}^{2}}
{4\mu_{o}^{2}}
+\frac{1}{6} \log \frac{m_{H^{+}}^{2}+2m_{H^{o}}^{2}+m_{A^{o}}^{2}}
{4\mu_{o}^{2}}\nonumber\\
&-&\frac{\cdw}{3} \log
\frac{2m_{H^{+}}^{2}+m_{H^{o}}^{2}+m_{A^{o}}^{2}}{4\mu_{o}^{2}}\,,\nonumber\\ 
g_{6}&=&\frac{\sw^4}{2}\log \frac{m_{H^{o}}^{2}+m_{H^{+}}^{2}}{2\mu_{o}^{2}}
+\frac{\sw^4}{2}\log \frac{m_{A^{o}}^{2}+m_{H^{+}}^{2}}{2\mu_{o}^{2}}
-\sw^2 \log \frac{m_{H^{+}}^{2}+m_{H^{o}}^{2}+m_{A^{o}}^{2}}{3\mu_{o}^{2}}\nonumber\\
&+&\sw^2\frac{\cdw}{2}\log 
\frac{2m_{H^{+}}^{2}+m_{A^{o}}^{2}}{3\mu_{o}^{2}}+
\sw^2\frac{\cdw}{2}\log 
\frac{2m_{H^{+}}^{2}+m_{H^{o}}^{2}}{3\mu_{o}^{2}}\nonumber\\
&+&\frac{\cdw^2}{6} \log 
\frac{3m_{H^{+}}^{2}+m_{H^{o}}^{2}}{4\mu_{o}^{2}}+\frac{\cdw^2}{6} \log 
\frac{3m_{H^{+}}^{2}+m_{A^{o}}^{2}}{4\mu_{o}^{2}}
-\frac{\cdw}{3} \log
\frac{2m_{H^{+}}^{2}+m_{H^{o}}^{2}+m_{A^{o}}^{2}}{4\mu_{o}^{2}}\nonumber\\ 
&+& \frac{1}{6} \log \frac{m_{H^{+}}^{2}+m_{H^{o}}^{2}+2m_{A^{o}}^{2}}
{4\mu_{o}^{2}}
+\frac{1}{6} \log \frac{m_{H^{+}}^{2}+2m_{H^{o}}^{2}+m_{A^{o}}^{2}}
{4\mu_{o}^{2}}\,,\nonumber\\
g_{7}&=&-\frac{2}{3}\log \frac{m_{H^{o}}^{2}+m_{H^{+}}^{2}}{2\mu_{o}^{2}}
-\frac{2}{3}\log \frac{m_{A^{o}}^{2}+m_{H^{+}}^{2}}{2\mu_{o}^{2}}\nonumber\\
g_{8}&=&-\log \frac{m_{H^{+}}^{2}}{\mu_{o}^{2}}-\frac{1}{2}
\log \frac{m_{H^{o}}^{2}}{\mu_{o}^{2}}-\frac{1}{2}
\log \frac{m_{A^{o}}^{2}}{\mu_{o}^{2}}+
\log \frac{2m_{H^{o}}^{2}+m_{H^{+}}^{2}}{3\mu_{o}^{2}}\nonumber\\
&+&\log \frac{2m_{H^{+}}^{2}+m_{H^{o}}^{2}}{3\mu_{o}^{2}}+
\log \frac{2m_{A^{o}}^{2}+m_{H^{+}}^{2}}{3\mu_{o}^{2}}+
\log \frac{2m_{H^{+}}^{2}+m_{A^{o}}^{2}}{3\mu_{o}^{2}}\nonumber\\
&-&\frac{2}{3}\log \frac{m_{H^{+}}^{2}+m_{H^{o}}^{2}}{2\mu_{o}^{2}}-
\frac{2}{3}\log \frac{m_{H^{+}}^{2}+m_{A^{o}}^{2}}{2\mu_{o}^{2}}\,.
\end{eqnarray} 
Notice that all these $g_k$ functions behave in the decoupling limit 
generically as,
\begin{equation}
g_{k}(m_{H^{+}},m_{H^{o}},m_{A^{o}})=
O \left ( \log \frac{m_{A^{o}}^2}{\mu_0^2}\right ) +
O \left ( \frac{\Delta m^2}{\Sigma m^2} \right )
\end{equation}
and the differences $\Delta m^2$ vanish in the present case of the
heavy MSSM Higgs sector in the asymptotic limit.
 
\vspace{0.5cm}

\begingroup\raggedright\endgroup

\end{document}